\documentclass[twocolumn,aps,showpacs,prr,preprintnumbers,longbibliography,superscriptaddress]{revtex4-2}

\usepackage{physics}
\usepackage{graphicx}
\usepackage{amsmath}
\usepackage[hidelinks]{hyperref}
\usepackage{dsfont}
\usepackage[yyyymmdd,hhmmss]{datetime}
\usepackage{tipa}

\begin{document}
\title{Encoding a one-dimensional topological gauge theory\\ in a Raman-coupled Bose-Einstein condensate}
\author{C. S. Chisholm}
\affiliation{ICFO - Institut de Ciencies Fotoniques, The Barcelona Institute of Science and Technology, 08860
Castelldefels (Barcelona), Spain}
\author{A. Fr\"{o}lian}
\affiliation{ICFO - Institut de Ciencies Fotoniques, The Barcelona Institute of Science and Technology, 08860
Castelldefels (Barcelona), Spain}
\author{E. Neri}
\affiliation{ICFO - Institut de Ciencies Fotoniques, The Barcelona Institute of Science and Technology, 08860
Castelldefels (Barcelona), Spain}
\author{R. Ramos}
\affiliation{ICFO - Institut de Ciencies Fotoniques, The Barcelona Institute of Science and Technology, 08860
Castelldefels (Barcelona), Spain}
\author{L. Tarruell}\email{Electronic address: leticia.tarruell@icfo.eu}
\affiliation{ICFO - Institut de Ciencies Fotoniques, The Barcelona Institute of Science and Technology, 08860
Castelldefels (Barcelona), Spain}
\affiliation{ICREA, Pg. Llu\'{i}s Companys 23, 08010 Barcelona, Spain}
\author{A. Celi}\email{Electronic address: alessio.celi@uab.cat}
\affiliation{Departament de F\'{i}sica, Universitat Aut\`{o}noma de Barcelona, E-08193 Bellaterra, Spain}
\date{\ddmmyyyydate\today}

\begin{abstract}
Topological gauge theories provide powerful effective descriptions of certain strongly correlated systems, a prime example being the Chern-Simons gauge theory of fractional quantum Hall states. Engineering topological gauge theories in controlled quantum systems is of both  conceptual and practical importance, as it would provide access to systems with exotic excitations such as anyons without the need for strong correlations. Here, we discuss a scheme to engineer the chiral BF theory, a minimal model of a topological gauge theory corresponding to a one-dimensional reduction of the Chern-Simons theory, with ultracold atoms. Using the local conservation laws of the theory, we encode its quantum Hamitonian into an ultracold quantum gas with chiral interactions. Building on a seminal proposal by Edmonds \emph{et al.} (Phys. Rev. Lett. \textbf{110}, 085301 (2013)), we show how to implement it in a Raman-coupled Bose-Einstein condensate with imbalanced scattering lengths, as we have recently realized experimentally (Fr\"{o}lian \emph{et al.,} Nature \textbf{608}, 293 (2022)). We discuss the properties of the chiral condensate from a gauge theory perspective, and assess the validity of the effective quantum description for accessible experimental parameters \emph{via} numerical simulations. Our approach lays the foundation for realizing topological gauge theories in higher dimensions with Bose-Einstein condensates.
\end{abstract}

\maketitle

\section{Introduction\label{Sec1}}
Ultracold quantum gases constitute an ideal realization of Feynman's vision of a quantum simulator \cite{Feynman1982}. Thanks to their excellent level of control, during the last two decades they have allowed the exploration of strongly-correlated condensed matter models which are hard to investigate with classical computing methods, such as fermionic superfluidity or the Hubbard model, as well as of weakly interacting systems in extreme parameter regimes which are out of reach for traditional solid-state platforms, e.g. enabling the realization of the celebrated Hofstadter and Haldane models \cite{Bloch2008, Esslinger2010, Bloch2012, Gross2017, Cooper2019}. The last years have witnessed increased efforts to extend this strategy to the quantum simulation of gauge theories \cite{Wiese2013, Zohar2016, Dalmonte2016, Banuls2020, Klco2022}. In this paper, we further extend this program to include topological gauge theories.

Gauge theories are key to our understanding of interactions between elementary particles as mediated by gauge bosons \cite{Cottingham2007,Altarelli2017,Peskin2018} and constitute powerful effective descriptions of the low-energy properties of strongly correlated condensed-matter systems \cite{Wen2004,Fradkin2013}.  As in the quantum simulation of condensed-matter physics, the motivation to emulate gauge theories with ultracold atoms is two-fold. On one hand, quantum simulators promise access to the equilibrium and out-of-equilibrium properties of strongly coupled lattice gauge theories that are very hard to determine by classical means, like the phase diagram of quantum chromodynamics at high density \cite{Fukushima2011} or the dynamics of the quark-gluon plasma \cite{Berges2021}. On the other hand, they provide a versatile platform for engineering the effective gauge theories that describe certain strongly correlated systems with exotic excitations in a weakly interacting and well controlled setting. These platforms may grant access to anyons or Majorana fermions without the need for strong interactions \cite{Weeks2007}.

Simulating gauge theories with ultracold atoms is, however, fundamentally more challenging than emulating Hubbard or spin models. To account for both the matter and gauge fields, multicomponent systems are required. Moreover, to ensure the gauge invariance of the theory, the matter and gauge fields must obey specific local conservation laws, i.e. local symmetry constraints that link them at every point of space and time. Enforcing such constrained dynamics on a quantum simulator is very demanding and has been the focus of intense theoretical research for more than a decade \cite{Buechler2005, Weimer2010, Zohar2011, Banerjee2012, Banerjee2013, Kasamatsu2013, Tagliacozzo2013b, Tagliacozzo2013a, Zohar2013, Stannigel2014, Glaetzle2014, Notarnicola2015, Zohar2015, Kasper2016, Cardarelli2017, Dutta2017, Muschik2017, Bender2018, Gazit2018, Lamm2018, Zache2018, Zhang2018, Barbiero2019, Lamm2019, Borla2020, Celi2020, Gonzalez2020, Kasper2020, Notarnicola2020, Armon2021, Halimeh2021a, Kebric2021, Paulson2021, Verresen2021}. Implementing the resulting protocols in the laboratory is even more challenging, which explains the comparatively small number of experimental realizations to date \cite{Banuls2020, Aidelsburger2021, Klco2022}.

The vast majority of experiments have been focused on the quantum simulation of gauge theories where the gauge fields are dynamical, i.e. they have propagating solutions even in the absence of matter. The prototypical example of this type of gauge theory is quantum electrodynamics (QED), which describes the coupling of electrons to photons. The local conservation law that ensures its Abelian U(1) gauge invariance is Gauss' law. In one dimension, however, no propagating gauge degrees of freedom are allowed. Indeed, in this case there are no transverse electric field components: the electric field is longitudinal and completely determined by the charges through Gauss' law, greatly simplifying its experimental implementation.

A discretized one-dimensional version of QED, the Schwinger model, is widely used as toy model for quantum simulation approaches. It has been implemented in trapped ion quantum computers \cite{Martinez2016, Kokail2019} by eliminating the gauge field through Gauss' law and explicitly encoding it on matter degrees of freedom \cite{Muschik2017}. A one-dimensional U(1) quantum-link model akin to QED, where the electric field is truncated to a spin of $1/2$, has been realized recently in an optical lattice system \cite{Yang2020, Zhou2021}. This experiment explicitly implements both gauge and matter degrees of freedom, and imposes Gauss' law through suitably designed energy penalties. Rydberg chain experiments \cite{Bernien2017} can be interpreted as an encoded version of the same model \cite{Surace2020}.  Building blocks of $\mathds{Z}_2$ \cite{Dai2017, Schweizer2019} and U(1) \cite{Klco2018, Mil2020} lattice gauge theories have also been demonstrated, paving the way towards the realization of the corresponding lattice gauge theories in extended quantum systems, and this approach has even been investigated using classical electric circuits \cite{Riechert2021}.

Interestingly, analogous dynamical gauge theories emerge naturally in the effective description of certain strongly correlated condensed matter systems, bridging the fields of high and low-energy physics \cite{Fradkin2013}. For instance, Abelian U(1) and $\mathds{Z}_2$ lattice gauge theories capture the behavior of certain quantum spin liquids \cite{Lacroix2011, Sachdev2016}, and Kitaev's toric code is a $\mathds{Z}_2$ lattice gauge theory \cite{Kitaev2006}. Minimal instances of the toric code have been implemented in quantum simulators using photonic, nuclear magnetic resonance, superconducting qubits, ultracold atoms, and trapped ion platforms \cite{Lu2009, Pachos2009, Feng2013, Zhong2016, Dai2017, Erhard2021}, and extended versions of both quantum spin liquids and  of the toric code have been recently demonstrated using Rydberg atom arrays \cite{Semeghini2021} and superconducting qubits \cite{Satzinger2021}.

Among the gauge theories emerging in condensed-matter physics, topological field theories play an important role \cite{Wen2004, Altland2006, Fradkin2013}. In this case, the gauge fields are unusual because they do not have propagating degrees of freedom in the absence of matter even beyond one spatial dimension. As a result, they can always be eliminated and the theory reformulated to involve only matter degrees of freedom. However, since this is done at the expense of introducing unconventional (and very often non-local) interactions between the particles, a gauge theory formulation for such effective theories is more natural and insightful.

A prime example of topological field theory is the Abelian U(1) Chern-Simons gauge theory which is commonly used as an effective low-energy and single-particle description of two-dimensional fractional quantum Hall states. It builds upon Wilczek's idea of \emph{flux attachment} \cite{Wilczek1982, Ezawa2008}. It consists of replacing the strongly interacting matter particles of the original system, which were subjected to a large external magnetic field perpendicular to the two-dimensional plane, with weakly interacting particles with an integer number of magnetic flux quanta attached to each of them. Chern-Simons theory is the field theory describing the coupling of these new weakly interacting particles to the vector potential that generates the flux tubes' magnetic field. The strength of the Chern-Simons description lies in the fact that the new composite particles (particle+attached flux tubes) provide a natural single-particle interpretation of the peculiar properties of fractional quantum Hall states. For instance, the fractional transverse conductance plateaus simply correspond to the filling of the composite particles' Landau levels, and the quasi-particles’ anyonic character emerges from the Aharonov-Bohm phases that are picked up upon exchange of flux tubes \cite{Ezawa2008}. Therefore, directly engineering the Chern-Simons theory in quantum simulators is a powerful approach to harness the physics of anyons in a well controlled and weakly interacting setting \cite{Weeks2007}.

The local conservation law of the Chern-Simons theory is the flux attachment condition, i.e. the fact that the Chern-Simons magnetic field is proportional to the matter density. Thus, the theory can be implemented in encoded form by realizing a density-dependent vector potential whose curl is proportional to the matter density. Conversely, a model with a matter-dependent gauge field corresponds to a topological gauge theory if and only if the dependence follows from a local conservation law. Recent experiments have realized density-dependent Peierls phases, i.e. density-dependent vector potentials in lattice systems, using ultracold atoms \cite{Clark2018, Gorg2019, Yao2022}, superconducting qubits \cite{Roushan2017}, and Rydberg atoms \cite{Lienhard2020}, but in these works the local conservation laws required to implement gauge invariant topological gauge theories were not enforced.

Here, we approach the problem of engineering a topological gauge theory from a different perspective. Requiring a precise matter-dependent gauge field is equivalent to requiring a precise form of the matter-matter interactions. Armed by this simple but powerful consideration, we show that a possible one-dimensional reduction of Chern-Simons theory in the continuum, the so-called chiral BF theory \cite{Rabello1995, Rabello1996, Aglietti1996, Jackiw1997, Griguolo1998}, can be realized by engineering chiral interactions in a Bose-Einstein condensate (BEC). We prove that the chiral BF theory reduces to a theory of bosonic matter fields with chiral interactions once the local conservation law is used to eliminate the gauge field. An analogous encoding was exploited in Refs. \cite{Martinez2016, Muschik2017, Kokail2019} to realize the Schwinger model, resulting in Coulomb interactions. Then, inspired by the seminal proposal of Ref. \cite{Edmonds2013}, we show that chiral interactions emerge in the low-energy quantum description of a Raman-coupled BEC with imbalanced interstate scattering lengths. The proposed implementation, which we very recently realized in Ref. \cite{Froelian2022}, ensures gauge invariance by construction and is directly implemented in the continuum. Since the formulation of Chern-Simons theory is considerably less involved in the continuum than in the lattice \cite{Frohlich1989,Kantor1991,Eliezer1992,Sun2015}, extending our experimental strategy to two-dimensional systems could allow for the realization of Chern-Simons theory, as has been proposed recently \cite{Valenti2020}. Thus, we expect the chiral BF theory to play a role model in the quantum simulation of topological gauge theories with ultracold atoms, similar to the one played by the Schwinger model in the quantum simulation of dynamical lattice gauge theories.

In order to provide the theoretical basis of the chiral BF theory, its emergence in Raman-coupled BECs, and discuss the validity of its implementation in realistic experimental conditions, this paper is organized as follows. In Section \ref{Sec2} we review the main properties of topological field theories using the Lagrangian formalism, emphasize the differences between topological and dynamical gauge theories, and derive the chiral BF model as a dimensional reduction of Chern-Simons theory supplemented with an additional chiral boson term \cite{Rabello1995, Rabello1996, Aglietti1996, Jackiw1997, Griguolo1998}. In Section \ref{Sec3}, we derive a consistent Hamiltonian for the chiral BF theory suitable for experimental realizations. It is obtained by applying the so-called Faddeev-Jackiw formalism to the chiral BF Lagrangian \cite{Faddeev1988,Jackiw1993}, which yields a Hamiltonian in second-quantized and encoded form, where the gauge degrees of freedom have been consistently eliminated in terms of the matter ones. Section \ref{Sec4} focuses on the implementation of this second-quantized encoded Hamiltonian in Raman-coupled BECs with imbalanced interactions, and presents the main result of this work: the microscopic derivation of the chiral BF theory as the effective quantum Hamiltonian of such systems. Section \ref{Sec5} analyzes the phenomenology of chiral BECs from a topological field theory perspective. In Section \ref{Sec6} we investigate the validity of the mapping to the chiral BF model for realistic experimental parameters, as realized in Ref. \cite{Froelian2022} with a $^{39}$K BEC, by means of numerical simulations. Finally, in Section \ref{Sec7} we draw the conclusions and perspectives opened by this work.

\section{Topological field theories and the chiral BF theory\label{Sec2}}

The chiral BF theory is a possible dimensional reduction of the U(1) Chern-Simons theory and constitutes one of the simplest examples of a topological field theory \cite{Rabello1995, Rabello1996, Aglietti1996, Jackiw1997, Griguolo1998}. As discussed in Section \ref{Sec1}, topological field theories are very particular types of gauge theories. They describe the coupling of matter fields to gauge fields that do not have propagating degrees of freedom in vacuum. Instead, their dynamics are linked to those of matter through the local symmetry constraint of the theory, which also ensures its gauge invariance. In this Section, we start by reviewing the main features of topological field theories and their differences with respect to dynamical gauge theories. We do so by comparing the U(1) Chern-Simons theory and electromagnetism. We then introduce the chiral BF theory, obtained by reducing Chern-Simons theory from two to one spatial dimensions and supplementing the resulting Lagrangian with a chiral boson term \cite{Aglietti1996, Jackiw1997, Griguolo1998}. We restrict ourselves to a system with open boundary conditions, as discussed in the literature. The interesting case of closed boundary conditions, where magnetic degrees of freedom appear even in one dimension, will be the subject of future work. Finally, we derive the equations of motion of the matter and gauge fields for the chiral BF theory, highlighting the analogies to the Chern-Simons case.

\emph{Maxwell theory.} Electromagnetism, i.e. Maxwell theory, is the best known example of a dynamical gauge theory. It is a U(1) Abelian field theory that describes a matter field $\Psi$, which we take to be non-relativistic and bosonic, minimally coupled to a U(1) gauge field $A_\mu$. Here and in the following, greek indices indicate the space-time coordinates $0,\dots,d$, with $0$ corresponding to the time component $t$ and $1,\dots,d$ (where $d$ can go up to $3$) corresponding to the spatial components $x,\,y,\,z$. The Lagrangian density of Maxwell theory is
\begin{equation}
{\cal L}_{\mathrm{M}} = -\frac 14 F_{\mu\nu} F^{\mu\nu} - A_\mu J^\mu +\mathcal{L}_{\mathrm{matter}}, \label{eq:Max}
\end{equation}
where
\begin{equation}
\mathcal{L}_{\mathrm{matter}}= i \Psi^* \dot \Psi +\frac {1}{2m}\Psi^*\nabla^2\Psi-V(n). \label{eq:L-matter}
\end{equation}
In these expressions, the time component of the current is simply the density, $J^0 = n = \Psi^*\Psi$,
the spatial components are given by ${\bf J} = \left[\Psi^* (\boldsymbol{\boldsymbol{\nabla}} - i {\bf A})\Psi - ((\boldsymbol{\nabla} + i {\bf A})\Psi^*)\Psi\right]/(2im)$, where $m$ is the mass of the matter particles, and $F_{\mu\nu}=\partial_{\mu} A_{\nu}-\partial_{\nu} A_{\mu}$ is the electromagnetic field tensor. Moreover, here and in the following repeated indices are summed, indices are raised and lowered with the mostly negative Minkowski metric, spatial vectors are denoted in bold and with the index up, the dot denotes the time derivative, e.g. $\dot{\Psi}\equiv \partial_0 {\Psi}$, and $V(n)$ are the matter-matter interactions, which depend only on the matter density $n$. We have also set the reduced Planck constant $\hbar$, the speed of light $c$, the vacuum permittivity $\epsilon_0$, the vacuum permeability $\mu_0$, and the electron charge $e$ to $1$.

To clarify the content of the theory, we follow Faddeev and Jackiw \cite{Faddeev1988,Jackiw1993}
and rewrite the Lagrangian density in first-order formalism, i.e. in a form analogous to the Legendre transform of a Hamiltonian density
\footnote{In first-order formalism, one partially replaces the time-derivative of the field with its conjugate momentum in order to get an expression that is at most linear in the time-derivative of the field. For example, the classical mechanics Lagrangian of a particle of position $q$ and velocity $\dot{q}$ subjected to a potential $V(q)$, $L=m\dot{q}^2/2-V(q)$, becomes in first-order formalism $L=p\dot{q}-[p^2/(2m)+V(q)]=p\dot{q}-H(q,p)$.}
\par\noindent\begin{multline}
{\cal L}_{\mathrm{M}} =  A^0\left(\mathbf{\boldsymbol{\nabla}} \cdot {\bf E} - n \right) -{\bf E}\cdot \dot {\bf A}  - \frac {1}2 \left( {\bf E}^2 + {\bf B}^2 \right) +  \\
+ i \Psi^* \dot \Psi +\frac 1{2m}\Psi^*\left(\mathbf{\boldsymbol{\nabla}} -i{\bf A}\right)^2\Psi-V(n),\label{eq:Max_1st_ord}
\end{multline}
where we have introduced the electric and magnetic fields $E^i =({\bf E})^i = -F^{0i} = -(\dot A^i + \partial_i A^0)$ and $B^i =({\bf B})^i = (\mathbf{\boldsymbol{\nabla}}\times {\bf A})^i = -\frac 12 \epsilon^{ijk} F^{jk}$. Here and in the following we use latin indices for spatial coordinates, and $\epsilon^{ijk}$ is the Levi-Civita symbol. In this form of the Lagrangian density the term $-{\bf E}\cdot \dot {\bf A}$ is the desired symplectic form for the gauge field that will allow us to rewrite the Lagrangian in canonical form (see next section), and we have omitted the total derivative $\mathbf{\boldsymbol{\nabla}}\left(A^0 {\bf E}\right)$ because it does not contribute to the action.

Equation \eqref{eq:Max_1st_ord} does not contain terms of the form $\partial_{\mu}A^0$, and therefore $A^0$ plays the role of a Lagrange multiplier. It enforces the local conservation law necessary to ensure gauge invariance,
\begin{equation}
\div\mathbf{E} = n,\label{eq_Gauss}
\end{equation}
which corresponds to Gauss' law.

The equations of motion for the matter and gauge fields are simply
\begin{eqnarray}
&&i\dot\Psi+\frac{1}{2m}
(\boldsymbol{\nabla}-i \mathbf{A})^2\Psi
-\frac{\mathrm{d}V(n)}{\mathrm{d}n}\,\Psi=0\cr
&&\div\mathbf{E} =n,\, \curl\mathbf{B}-\dot{\mathbf{E}}=\mathbf{J}\cr
&&\div\mathbf{B} =0,\,\curl\mathbf{E}+\dot{\mathbf{B}}=0.
\label{eq:Max-EoM}
\end{eqnarray}
For the gauge field we recover Maxwell's equations. They show that, in the absence of matter, only the transverse (divergenceless) part of the gauge field has propagating solutions: the electromagnetic waves. For $d=2$ or $3$, the gauge field is thus \emph{dynamical}, making electromagnetism a \emph{dynamical gauge theory}. In one spatial dimension $d=1$, however, no transverse electric field nor magnetic fields appear. Thus, in $d=1$ the gauge field has no dynamics in vacuum.

\emph{Chern-Simons theory.} The prototypical example of a topological field theory, Chern-Simons theory, is also a U(1) Abelian gauge theory. It describes a matter field $\Psi$ evolving in two spatial dimensions $d=2$ and minimally coupled to a gauge field $A_\mu$ that is ``internal'', i.e. self-generated by the system. If no additional external fields are applied, its Lagrangian density reads
\begin{equation}
\mathcal{L}_{\mathrm{CS}} =\frac{1}{4\kappa}\epsilon^{\mu\nu\rho}A_{\mu}{F}_{\nu\rho}- A_\mu J^\mu +\mathcal{L}_{\mathrm{matter}}\label{eq:CS},
\end{equation}
where $\mu=0,1,2$ and the coefficient $\kappa$ of the gauge field term is the so-called Chern-Simons level. Its value indicates the number of flux tubes that are attached to each matter particle, and the particular fractional quantum Hall state described. For the case of a non-relativistic and bosonic matter field considered in this work, see Eq. \eqref{eq:L-matter}, $\mathcal{L}_{\mathrm{CS}}$ is also known as the Jackiw-Pi model \cite{Jackiw1990a, Jackiw1990b}.

As we did above for electromagnetism, we separate the $A^0$ term from the Lagrangian. Moreover, we introduce ${\bf A}$ and ${\bf B}=B\mathbf{e}_3$, where the magnetic field now has only one component that is perpendicular to the system's plane. We then obtain the Lagrangian density
\par\noindent\begin{multline}
\mathcal{L}_{\mathrm{CS}} =
-A^0\left(\frac{B}{\kappa}+n\right)
-\frac{1}{2\kappa}{\bf A}\times {\bf \dot A}\\
+ i \Psi^* \dot \Psi +\frac 1{2m}\Psi^*\left(\mathbf{\boldsymbol{\nabla}} -i{\bf A}\right)^2\Psi-V(n)\label{eq:CS2}.
\end{multline}

In this case, the local conservation law of the theory, i.e. the equivalent to Gauss' law for electromagnetism, is
\begin{equation}
B= -\kappa n, \label{eq:FluxAttachment}
\end{equation}
a relation that is known as \emph{flux attachment} because it highlights the fact that $B$ is given by the flux tubes attached to the matter particles.

In the absence of matter, the equation of motion for the gauge field is $F_{\mu\nu}=0$, which is locally trivial. Thus, the existence of solutions for $A_\mu$ that are globally non trivial depends on the topology of the space. The situation is very different from that encountered in electromagnetism, where $\partial_{\mu}F^{\mu\nu}=0$ (plus $\partial_\mu F_{\nu\rho} + \partial_\nu F_{\rho\mu} + \partial_\rho F_{\mu\nu}=0$) \cite{Jackson1999classical} and the gauge field has propagating solutions in vacuum: the electromagnetic waves.

In the presence of matter, the Chern-Simons field acquires non-trivial dynamics. This can be seen by explicitly writing the equations of motion for the matter and gauge fields
\begin{eqnarray}
&&i\dot\Psi+\frac 1{2m}
(\mathbf{\boldsymbol{\nabla}}  - i {\bf A})^2\Psi
-\frac{\mathrm{d}V(n)}{\mathrm{d}n}\,\Psi=0\cr
&&{\bf E}- \kappa \, \mathbf{e}_3 \times {\bf J}=0\cr
&&B+\kappa n=0. \label{eq:CS-EoM}
\end{eqnarray}

The equation for the electric field, combined with the continuity equation for the matter field $\partial_0 n +\div \mathbf{J}=0$, is equivalent to the equation of motion for the magnetic field. Thus, one sees that the flux attachment condition is simultaneously the equation of motion of the gauge field and the local conservation law of the theory. This is a property of gauge theories that do not have independent propagating degrees of freedom for the gauge field, and is analogous to what happens for electrodynamics in one dimension.

\emph{Chiral BF theory.} In the 90's, Refs. \cite{Rabello1995, Rabello1996} introduced a possible reduction of Chern-Simons theory from two to one spatial dimension that could conserve its main topological features.
This theory is obtained from the Jackiw-Pi model of Eq. \eqref{eq:CS} by removing the dependence in the $y$ spatial coordinate and setting $A_2=m \mathcal{B}$. Here $\mathcal{B}$ is a new bosonic scalar field that depends only on $x$ and $t$. Making the additional replacements $A^0 \to A^0- m \mathcal{B}^2/2$ and $\kappa \to m\kappa$, the Lagrangian density becomes
\begin{equation}
\mathcal{L}_{\mathrm{BF}} =\frac{1}{2\kappa}\mathcal{B}\epsilon^{\mu\nu}F_{\mu\nu}-A_\mu J^\mu +\mathcal{L}_{\mathrm{matter}}\label{eq:BF},
\end{equation}
with $\mu=0,1$.
Due to the form of the gauge field term (which replaces the Chern-Simons form $\epsilon^{\mu\nu\rho} A_{\mu} F_{\nu\rho}$, see Eq. \eqref{eq:CS}) this model receives the name of BF theory.
Because it does not include any derivative of $\mathcal{B}$, the equation of motion of the gauge field  is $F_{\mu\nu}=0$ even in the presence of matter.  This means that in this one-dimensional problem the $\mathcal{B}$ and $A_{\mu}$ fields can be completely decoupled from matter and eliminated by a phase redefinition of $\Psi$. The matter-matter interactions would then be solely determined by $V$, making the theory trivial. Refs. \cite{Rabello1995,Rabello1996,Aglietti1996,Jackiw1997,Griguolo1998} therefore add to the Lagrangian density a kinetic term for the gauge field of the self-dual form $\dot{\mathcal{B}}\mathcal{B}'$ \cite{Floreanini1987}. Here and in the following, the prime indicates the spatial derivative, i.e. $\mathcal{B}'\equiv \partial_1 \mathcal{B}$. This term explicitly breaks Galilean invariance and is the simplest non-relativistic combination that endows the system with chiral dynamics, making it reproduce the behavior of one edge of the original 2D system. Its chirality can be selected through the sign of $\lambda$. For this reason, it is called the chiral boson term.

The resulting model is the so-called chiral BF theory, whose Lagrangian density reads
\begin{equation}
\mathcal{L}_{\mathrm{cBF}} =\frac{1}{2\kappa}\mathcal{B}\epsilon^{\mu\nu}F_{\mu\nu}+\frac{\lambda}{2\kappa^2} \dot{\mathcal{B}}\mathcal{B}'- A_\mu J^\mu+\mathcal{L}_{\mathrm{matter}},\label{eq:BF}
\end{equation}
where $\lambda$ determines the strength of the chiral boson term. Specifying the electromagnetic field tensor in components (and up to total spatial derivatives), we have
\par\noindent\begin{multline}
\mathcal{L}_{\mathrm{cBF}} =
A^0\left(\frac{\mathcal{B}'}{\kappa}- n\right)
-\frac{\mathcal{B}}{\kappa}\dot A + \frac{\lambda}{2 \kappa^2} \dot{\mathcal{B}}\mathcal{B}'
\\+i\Psi^*\dot\Psi
+\frac{1}{2m}\Psi^*(\partial_1-i A)^2\Psi
-V(n),
\label{eq:BF_2}
\end{multline}
where the vector potential $A$ is now a scalar as in $d=1$ it has only one component.

From Eq. \eqref{eq:BF_2} we readily identify the local conservation law of the chiral BF theory
\begin{equation}
{\cal B}'=\kappa n, \label{eq:Bprime}
\end{equation}
which is the equivalent to the flux attachment condition in Chern-Simons theory or Gauss' law in electromagnetism.

The equations of motion for the matter and gauge fields read
\begin{eqnarray}
&&i\dot\Psi+\frac{1}{2m}
(\partial_1-i A)^2\Psi
-\frac{\mathrm{d}V(n)}{\mathrm{d}n}\,\Psi=0\cr
&&\dot{\mathcal{B}}+\kappa J=0,\,\mathcal{B}'-\kappa n=0\cr
&&E -\frac{\lambda}{\kappa} \dot{\mathcal{B}}'=0
\label{eq:BF-EoM}
\end{eqnarray}
where the electric field $E$ and the spatial part of the current $J=\left[\Psi^*(\partial_1 - i A)\Psi
-((\partial_1+i A) \Psi^*)\Psi\right]/(2im)$ are indicated as scalars because in $d=1$ they have only one component.

As in the Chern-Simons case, the first of these equations describes the motion of matter coupled to a gauge potential $A$,
and the two equations in the second line can be combined to yield the continuity equation for the matter field  $\partial_0 n+ \partial_1 J=0$.
The last expression is the equation of motion of the ``electromagnetic'' field tensor, which can be rewritten as $F_{01}=E =\lambda \dot{n}$.
Thus, we see that at the classical level the local conservation law ${\cal B}'=\kappa n$ is equivalent to $E=\lambda \dot{n}$ or $A=-\lambda n+\partial_1\Lambda$, where $\Lambda$ is an arbitrary function that depends only on the spatial coordinate.

In conclusion, we see that despite its simplicity the chiral BF theory already contains the main features of a topological field theory. It can thus be used as toy model to benchmark the quantum simulation of topological field theories with quantum gases. However, to achieve this we need a Hamiltonian formulation of the theory.

\section{Derivation of an encoded Hamiltonian\label{Sec3}}

In this Section, we tackle the problem of deriving a quantum Hamiltonian for the chiral BF theory that is amenable to quantum simulation with ultracold atoms. Since the theory is subjected to the local constraint Eq. \eqref{eq:Bprime} and the relations between $\dot A$ and $\dot {\cal B}$ and the conjugate momenta of $A$ and $\cal{B}$, $-{\cal B}/\kappa$ and $\lambda {\cal B}'/(2\kappa^2)$ respectively, are singular, i.e. they cannot be inverted, it is not straightforward to perform the Legendre transform of Eq. \eqref{eq:BF_2}.
To deal with the constrained system, we apply the first-order approach due to Faddeev and Jackiw \cite{Faddeev1988,Jackiw1993}. It allows one to separate the dynamics from the local conservation laws by progressively eliminating the dependent fields at the level of the Lagrangian. In this way, we avoid the complex Dirac treatment of constraints \cite{Dirac1950,Henneaux2020}. We end up with a second-quantized Hamiltonian involving only the physical degrees of freedom of the system, and where the matter-dependent gauge degrees of freedom have been eliminated using the local conservation law. That is, a Hamiltonian that has an {\it encoded} form similar to the one exploited to simulate the Schwinger model \cite{Martinez2016, Muschik2017}.

Since the Faddeev-Jackiw approach is not commonly used in quantum simulation, we first apply it to Maxwell theory and show the emergence of Gauss' law and of Coulomb's Hamiltonian. We then perform an analogous consistent elimination for the chiral BF theory, obtaining the local conservation law Eq. \eqref{eq:Bprime} already derived in Section \ref{Sec2} and the encoded Hamiltonian.

In both cases, the elimination of the matter-dependent gauge field produces a non-trivial interaction for the matter field: an infinite-range Coulomb interaction in the case of electromagnetism, and an anomalous chiral interaction term in the chiral BF theory. The implementation of the encoded Hamiltonian in an experimental system reduces then to the engineering of the corresponding emerging interaction term.

\emph{Maxwell theory.} The starting point of the Faddeev-Jackiw approach is the Lagrangian of the system in first-order form, i.e. Eq. \eqref{eq:Max_1st_ord} in the case of Maxwell theory.
There, the electric field of the system can be decomposed in longitudinal (i.e. stemming from the gradient of a function) and transverse (i.e. divergenceless) parts ${\bf E}={\bf E}_{\mathrm{L}} +{\bf E}_{\mathrm{T}}$, and Gauss' law can then be used to express the longitudinal part in terms of the matter field ${\bf E}_{\mathrm{L}}=\mathbf{\boldsymbol{\nabla}}(\nabla^{-2} n)$. Here $\nabla^{-2} n$ is simply minus the electrostatic potential. By substituting in the Lagrangian and decomposing also the vector potential in longitudinal and transverse parts, ${\bf A}={\bf A}_{\mathrm{L}} +{\bf A}_{\mathrm{T}}$, we find (up to total derivatives) \footnote{We have exploited that the inverse Laplacian can be integrated by parts like the other differential operators, $\int\mathrm{d}^3\,\mathbf{r} f \boldsymbol{\nabla}^{-2}g= \int \mathrm{d}^3\mathbf{r}\, \boldsymbol{\nabla}^{-2} f \, g$.}
\par\noindent\begin{multline}
\mathcal{L}_{\mathrm{M}} = -{\bf E}_{\mathrm{T}}\cdot \dot{{\bf A}}_{\mathrm{T}}  + i \Psi^* \dot{\Psi} + n\nabla^{-2}(\div {\dot {\bf A}}_{\mathrm{L}})\\
- \frac 12 \left( {\bf E}_{\mathrm{T}}^2 - n \nabla^2 n +  {\bf B}^2 \right)  + \frac {1}{2m}\Psi^*\left(\mathbf{\boldsymbol{\nabla}} -i{\bf A}\right)^2\Psi-V(n).\label{eq:Max_1st_ord2}
\end{multline}
This expression shows that the transverse components of the gauge field and the matter field are the dynamical degrees of freedom of the theory: the pairs $({\bf A}_{\mathrm{T}},-{\bf E}_{\mathrm{T}})$ and $(\Psi,i \Psi^*)$ are equivalent to position and momentum of a mechanical system, $(q,p)$. While the first two terms in Eq. \eqref{eq:Max_1st_ord2} have the canonical form $p \dot q$, the third term does not because there is no momentum field conjugated to ${\bf A}_{\mathrm{L}}$. This simply reflects the fact that the longitudinal component of the gauge field is not physical. The key point of the Faddeev-Jackiw treatment is that such non-canonical terms can always be removed by field redefinitions.

Indeed, if we redefine the matter field through the gauge transformation $\Psi =\exp[i \nabla^{-2}(\div \mathbf{A}_\mathrm{L})]\phi$, we obtain
\begin{eqnarray}
i \Psi^* \dot{\Psi}+ n\nabla^{-2}(\div \dot {\bf A}_{\mathrm{L}})&=& i \phi^* \dot{\phi}\mbox{\,\,\,\,\,and}\cr
\frac {1}{2m}\Psi^*\left(\mathbf{\boldsymbol{\nabla}} -i{\bf A}_{\mathrm{L}}\right)^2\Psi &=& \frac {1}{2m}\phi^*\nabla^2\phi.
\end{eqnarray}
In the second expression, we have used that by construction ${\bf A}_{\mathrm{L}}=\mathbf{\boldsymbol{\nabla}} f_{\mathrm{L}}$ for some function $f_{\mathrm{L}}$. Thus, $\div \mathbf{A}_{\mathrm{L}}=\nabla^2 f_{\mathrm{L}}$ and $f_{\mathrm{L}}=\nabla^{-2}(\div \mathbf{A}_{\mathrm{L}})$, which implies $\boldsymbol{\nabla}(\nabla^{-2} \div \mathbf{A}_{\mathrm{L}}) =\mathbf{A}_{\mathrm{L}}$.

Substituting this into Eq. \eqref{eq:Max_1st_ord2}, the Lagrangian density takes the canonical form
\begin{equation}
\mathcal{L}_{\mathrm{M}} =-{\bf E}_{\mathrm{T}}\cdot \dot {\bf A}_{\mathrm{T}}  + i \phi^* \dot \phi - {\cal H}^{\mathrm{enc}}_{\mathrm{C}}
\end{equation}
and we can read off the encoded Hamiltonian density
\begin{equation}
{\cal H}^{\mathrm{enc}}_{\mathrm{C}} = \frac 12 \left( {\bf E}_{\mathrm{T}}^2 - n \nabla^{-2} n +  {\bf B}^2 \right) -\frac {1}{2m}\phi^*\left(\mathbf{\boldsymbol{\nabla}} -i{\bf A}_{\mathrm{T}}\right)^2\phi,
\end{equation}
which is Coulomb's Hamiltonian. Note that this result has been derived without assuming any gauge choice.

The derivation above is valid also when we restrict ourselves to one or two spatial dimensions. In $d=1$, there is no transverse component and the Hamiltonian simply contains the kinetic term of matter and the Coulomb interaction, which is of infinite range as $E_{\mathrm{L}} =\int \mathrm{d}x\, n$ and thus $-n\nabla^{-2} n= \left(\int \mathrm{d}x\, n\right)^2 =E_{\mathrm{L}}^2$. It has precisely the same form as the interaction term in the lattice Hamiltonian of the Schwinger model after encoding \cite{Muschik2017}, where the integral of the density is replaced by the sum of the charges (represented as spins by a Jordan-Wigner transformation) over the sites of the lattice. It is the experimental implementation of such an exotic interaction term that makes the realization of the encoded Hamiltonian difficult, a challenge that was successfully tackled with trapped ions \cite{Martinez2016, Kokail2019}. In $d>1$, encoded formulations of Maxwell's lattice gauge theory Hamiltonians that exploit the electromagnetic duality have been proposed in Refs. \cite{Kaplan2020, Bender2020, Haase2021, Paulson2021}, while link model dual formulations have been investigated in Refs. \cite{Celi2020,Banerjee2021}.

\emph{Chiral BF theory.} We now apply the Faddeev-Jackiw procedure to the Lagrangian of the chiral BF model Eq. \eqref{eq:BF} in order to derive the corresponding encoded Hamiltonian.
Since only the matter field enters canonically in the Lagrangian density, this is the dynamical field (as the chiral BF equations of motion Eq. \eqref{eq:BF-EoM} also indicate).

In order to consistently formulate the theory in Hamiltonian form, separating the dynamics from the constraints, we start by redefining the matter field as
\par\noindent\begin{multline}
\Psi = \text{exp}\left[i\left( \int_{x_0}^{x} {\rm d}\xi \,A(\xi,t)
- \int_{t_0}^{t} {\rm d}t' A^0(x_0,t')\right.\right.
\\ \left. \left.+ \frac \lambda{2 \kappa} {\cal B}(x_0,t)\right) \right] \psi, \label{eq:BF-gaugeT-1}
\end{multline}
where $x_0$ and $t_0$ are specific values of the space and time coordinates. This gauge transformation, which is similar to the one performed in Ref. \cite{Griguolo1998} on the equations of motion, allows us to rewrite
the Lagrangian Eq. \eqref{eq:BF_2} (up to total derivatives) in the form
\par\noindent\begin{multline}
{\cal L_{\mathrm{cBF}}} = \tilde  A^{0}\left(\frac {{\cal B}'}\kappa -   n\right)
- \frac{\lambda}{2\kappa} \dot {\cal B}(x_0,t) n
\\+\frac{\lambda}{2  \kappa^2} \dot{\cal B} {\cal B}'
+ i   \psi^*\dot \psi
+ \frac{1}{2m}\psi^* \psi'' - V(n),\label{eq:cBF}
\end{multline}
with the Lagrange multiplier $\tilde A^0 \equiv -\int_{x_0}^x {\rm d}\xi \, F_{01}(\xi,t)$. We thus see that the effect of the gauge transformation Eq. \eqref{eq:BF-gaugeT-1} is to separate the dynamical part of the gauge field, proportional to the bosonic gauge field $\cal{B}$, from the “trivial” part represented by the Lagrange multiplier $\tilde{A}_0$. The latter imposes the constraint $\mathcal{B}'= \kappa n$  or, equivalently,
\begin{equation}
{\cal B}(x,t) =  \kappa \int_{x_0}^x {\rm d}\xi\, n(\xi,t) + {\cal B}(x_0,t),\label{eq:BF-B}
\end{equation}
which as we explained in Section \ref{Sec2}, is the local conservation law of the chiral BF theory.
To derive the encoded Hamiltonian, we substitute the symmetry constraint Eq. \eqref{eq:BF-B} into Eq. \eqref{eq:cBF}, obtaining a Lagrangian where the gauge field has been completely eliminated and only the matter field appears
\par\noindent\begin{multline}
{\cal L_{\mathrm{cBF}}} =
 i  \, \psi^*\left(\partial_0 -i \frac{\lambda}2 \int_{x_0}^x{\rm d}\xi\, \dot{n}(\xi,t)\right)\psi
\\+ \frac{1}{2m} \psi^*\psi'' - V(n).\label{eq:cBF_2}
\end{multline}

However, the resulting expression is non-local, and thus not easily amenable to quantum simulation. Moreover, the canonical quantization of Eq. \eqref{eq:cBF_2} would lead to a quantum field that is not bosonic. This was to be expected, since in two spatial dimensions Chern-Simons theory is a field theory for anyons, and its one-dimensional reduction, the chiral BF theory, was originally constructed as a possible model for anyons in a line \cite{Rabello1995, Rabello1996}. Indeed, although in the original papers the quantum mechanical model proposed as microscopic realization of the theory was not correct \cite{Aglietti1996, Jackiw1997, Griguolo1998}, the chiral BF theory has been subsequently shown to be the field theory corresponding to the Kundu model, a well-defined microscopic model for linear anyons, in the regime of vanishing contact interactions \cite{Kundu1999}. In fact, as shown in Refs. \cite{Aglietti1996, Jackiw1997, Griguolo1998}, adding contact interactions does not alter the commutation relation of linear anyons, as it gives contact interactions for the anyonic field.

Our final step to derive an encoded Hamiltonian suitable for the quantum simulation of the chiral BF theory is to remove the non-locality of Eq. \eqref{eq:cBF_2} by performing the Jordan-Wigner transformation

\begin{equation}
\psi = \text{exp}\left[i\frac {\lambda}2 \int_{x_0}^{x} {\rm d}\xi \,n(\xi,t)\right] \phi \label{eq:cBF-psi},
\end{equation}
after which the matter field $\phi$ is again bosonic. This yields the local Lagrangian density
\begin{equation}
{\cal L_{\mathrm{cBF}}^{\mathrm{enc}}} =
 i  \, \phi^*\dot \phi
- {{\cal H}_{\mathrm{cBF}}^{\mathrm{enc}}}
\label{eq:cBF-L-enc}
\end{equation}
from which (up to total derivatives) we read off a Hamiltonian density that is canonical and can be quantized
\begin{equation}
{{\cal H}_{\mathrm{cBF}}^{\mathrm{enc}}} = -\frac{1}{2m} \phi^*\phi''+ \tilde V (n) + \frac{\lambda}2 J\,n
\label{eq:cBF-H-enc}.
\end{equation}
Here $J=\left(\phi^*\phi'-\phi^{*\prime}\phi\right)/(2im)$ is the spatial current for the non-relativistic boson $\phi$ not coupled to any gauge field and $\tilde V$ includes now a positive cubic term, $\tilde V(n) = V(n) + \lambda^2n^3/(8 m)$. Thus, the encoded model corresponds to that of a bosonic field with effective chiral and three-body interactions besides the original ones
\begin{equation}
\mathcal{H}_{\mathrm{int}}= V(n)+\mathcal{H}_{\mathrm{int}}^{\mathrm{chiral}}+\mathcal{H}_{\mathrm{int}}^{\mathrm{3B}}=V(n)+\frac{\lambda}2 J\,n + \frac{\lambda^2}{8 m}n^3.
\label{eq:cBF-H-enc-int}
\end{equation}

For small values of $\lambda$, the three-body term $\mathcal{H}_{\mathrm{int}}^{\mathrm{3B}}\propto \lambda^2$ becomes negligible, and the encoded model reduces to that of a bosonic field with additional chiral interactions $\mathcal{H}_{\mathrm{int}}^{\mathrm{chiral}}\propto \lambda$. Therefore, and similarly to the Maxwell case, the elimination of the gauge field through the local conservation law produces an interaction in the encoded Hamiltonian, which in the case of the chiral BF theory is not Coulomb-like but dominantly of current-density form.
This interaction is chiral: it breaks Galilean invariance as the chiral BF theory before the encoding does. For localized wavepackets, it is equivalent to a density-density term with a coupling that changes with the velocity of the wavepacket itself (see Section \ref{Sec5} for further details). Conversely, a non-relativistic bosonic system displaying a current-density interaction is described by a chiral BF theory, i.e. we can measure all the observables of the latter in the former.

The encoded Hamiltonian Eq. \eqref{eq:cBF-H-enc} provides a simple prescription for obtaining the quantum chiral BF theory, since it is sufficient to normal order it to have a second-quantized Hamiltonian operator \cite{Jackiw1997, Griguolo1998}
\begin{equation}
{{\cal \hat{H}}_{\mathrm{cBF}}^{\mathrm{enc}}} = -\frac{1}{2m} \hat{\phi}^{\dagger}\hat{\phi}''  + :\tilde V (\hat{n}):+ \frac{\lambda}2 :\hat{J}\,\hat{n}:
\label{eq:cBF-H-enc-q}
\end{equation}
Through the local constraint, we can then determine the expectation value of all of the observables. For instance, the expectation value of the electric field of the chiral BF theory before encoding can be obtained by measuring the time derivative of the expectation value of the density $\langle \hat{E}\rangle =\lambda \langle \dot{\hat{n}}\rangle$. Note that an analogous strategy was used to determine the electric field when simulating the encoded Schwinger model with trapped ions \cite{Martinez2016, Muschik2017}.

Coincidentally, Eq. \eqref{eq:cBF-H-enc} can be rewritten as
\begin{equation}
{{\cal H}_{\mathrm{cBF}}^{\mathrm{enc}}} = -\frac{1}{2m} \phi^*\left(\partial_1 + i \frac{\lambda}2 n\right)^2\phi + V(n)
\label{eq:cBF-H-enc-An}
\end{equation}
and interpreted as a matter field $\phi$ minimally coupled to a density-dependent vector potential ${\cal A} = -\lambda n/2$ \cite{Aglietti1996, Jackiw1997, Griguolo1998}. It was precisely this property of the encoded theory that inspired the first theoretical proposal for its implementation in a Raman-coupled BEC with imbalanced interactions \cite{Edmonds2013}. Note however that ${\cal A}$  is not the gauge potential of the chiral BF theory before encoding. This can be readily seen from the classical equations of motion of the theory Eq. \eqref{eq:BF-EoM}, from which (for a particular gauge choice) the chiral BF gauge potential can be written as $A=-\lambda n$. It differs from the definition of $\cal{A}$ by a factor of $2$.

\section{Experimental implementation in a Raman-coupled BEC\label{Sec4}}

The local Hamiltonian for the chiral BF theory in encoded form involves only a bosonic matter field and local interactions. This makes it amenable to quantum simulation with Bose-Einstein condensates, as originally put forward by Edmonds \emph{et al.} \cite{Edmonds2013}. In the first part of this Section, we briefly review the key results of this seminal theoretical proposal that, building upon the density-dependent vector potential interpretation of the encoded chiral BF Hamiltonian presented in Section \ref{Sec3}, maps a Raman-coupled BEC into the classical version of the encoded chiral BF theory. This is achieved using a position-space approach valid in the large Raman coupling regime. In the second part of this Section, we present an alternative derivation of the mapping to the encoded chiral BF theory. Based on a microscopic momentum-space picture, it highlights the chiral interaction interpretation of the encoded model and encompasses as well its quantum regime. We also exploit this momentum-space approach to show that under specific conditions, the mapping between a Raman-coupled BEC and the chiral BF model can be extended to the regime of moderate Raman coupling strength, which is experimentally more accessible \cite{Froelian2022}. From this Section to the end of the paper, we restore the usual notation for the time and space coordinates $\mu=0\rightarrow t$ and $\mu=1\rightarrow x$, and introduce dimensionless units in terms of the Raman-coupling parameters following conventions usual in the field \cite{Spielman2009}.

\begin{figure}[t]
\includegraphics{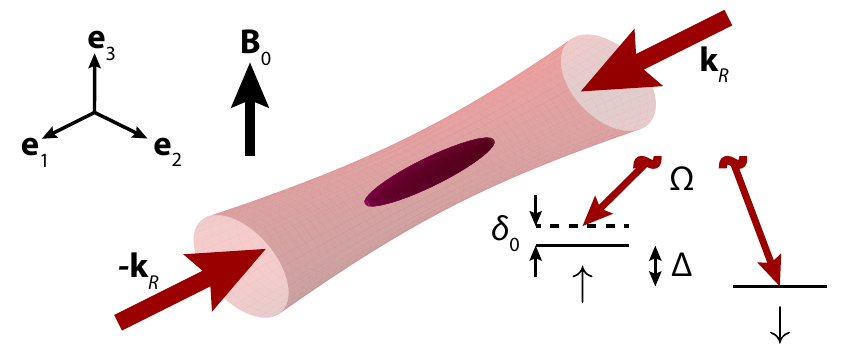}
\caption{\label{fig:Fig1} Experimental scheme. A Bose-Einstein condensate with two internal states $\uparrow$ and $\downarrow$ (purple ellipsoid) is transversally confined in an optical waveguide (light red cylinder). The two states have an energy difference $\Delta$ and are coupled by two Raman laser beams (red arrows) counter-propagating along the $x$ direction, with single-photon wavevectors $\pm k_R\mathbf{e}_1$, and two-photon Rabi frequency $\Omega$ and detuning $\delta_0$. An external magnetic field $\mathbf{B}_0$ is used to control the interatomic interactions.}
\end{figure}

\emph{Vector potential real-space derivation.}
The theoretical work of Edmonds \emph{et al.} proposes a scheme to engineer a bosonic field minimally coupled to a density-dependent vector potential and described by Eq. \eqref{eq:cBF-H-enc-An} with a Bose-Einstein condensate \cite{Edmonds2013}. Specifically, a vector potential $\cal A$ with the required linear density dependence is obtained by combining two ingredients:
\begin{enumerate}
\item [(i)] an optical coupling between two internal atomic states $\uparrow$ and $\downarrow$. Induced by two laser beams in Raman configuration (with two-photon Rabi frequency $\Omega$ and single-particle detuning $\delta_0$, see Fig. \ref{fig:Fig1}), it yields a synthetic vector potential that scales linearly with the detuning of the coupling field \cite{Dalibard2011};
\item [(ii)] differential intrastate interactions of the two states involved, given by the coupling constants $g_{\uparrow\uparrow}$ and $g_{\downarrow\downarrow}$, which makes this detuning density-dependent due to the differential mean-field shift of the transition.
\end{enumerate}
A BEC subjected to such a density-dependent synthetic vector potential is described by the encoded version of the chiral BF theory, and constitutes an ideal experimental platform to explore its properties.

Since the scheme has been reviewed in detail in previous works \cite{Edmonds2013, Goldman2014}, we only summarize the key concepts here. The mapping to the chiral BF theory is obtained under the adiabatic approximation for the atom-light coupling in position space. In this regime, the strength of the optical coupling $\Omega$ between the two atomic states is the dominant energy scale of the problem. Thus, the atoms adiabatically follow the eigenstates of the combined atom-photon system, which are position-dependent dressed states. This leads to a synthetic vector potential of “geometrical” origin \cite{Dalibard2011}. The interatomic interactions between the atoms are then added to the single-particle treatment of the problem within a mean-field approximation, and their effect on the non-interacting dressed states is included perturbatively. Thus, the chiral BF model of Eq. \eqref{eq:cBF-H-enc-An} is implemented at the mean-field level, i.e. as a classical field theory. Ref. \cite{Edmonds2013} shows as well that the dynamics of the atoms are described by an effective single-component extended Gross-Pitaevskii equation (eGPE) with an additional non-linear current term. It stems from the density-dependent vector potential $\cal A$, and for $\delta_0=0$ reads ${\cal A}=-\lambda n/2$, with $\lambda\propto (g_{\downarrow\downarrow}- g_{\uparrow\uparrow})/\Omega$.

Despite its appeal, this theoretical proposal is challenging to implement experimentally. First, it requires a bosonic atomic species with two internal states of sufficiently different scattering lengths in order to ensure that their differential mean-field energy is a sizeable fraction of the Rabi frequency, i.e. that $\lambda$ is large. This imposes the use of bosonic atoms with suitable Feshbach resonances, such as the alkali species lithium, potassium or cesium, or the lanthanide atoms dysprosium or erbium. Since the large dipolar interactions of the latter yield interaction terms that go beyond the chiral BF Hamiltonian of Eq. \eqref{eq:cBF-H-enc}, we focus on the alkali case. There, $^{39}$K provides two suitable magnetic field regions where $g_{\uparrow\uparrow}>0$, $g_{\downarrow\downarrow}>0$, the two intrastate scattering lengths are sufficiently imbalanced, and their magnitude is such that the mean-field description of the BEC remains valid \cite{Cheiney2018, Froelian2022}. In contrast, fulfilling all conditions simultaneously with $^{7}$Li or $^{133}$Cs is complicated. Second, the two states must be coupled using two laser beams of recoil energy $E_R$ exploiting a two-photon Raman scheme and, to reach the regime of validity of the position-space approach, a Rabi frequency on the order of $100E_R$ is needed \cite{Spielman2009}. Due to inelastic photon scattering from the Raman beams, this value is experimentally unrealistic for $^{39}$K, for which the lifetime of the BEC at $\Omega=100E_R/\hbar$ is below $2$ ms \cite{Wei2013}. To implement the chiral BF theory experimentally, it is thus important to relax the requirements on the Raman coupling strength.

\emph{Chiral interactions' momentum-space derivation.} At the single-particle level, it is well known that analyzing Raman-coupled systems in momentum space enables the identification of a synthetic vector potential proportional to the two-photon Raman detuning for much smaller values of the coupling strength than in the position-space approach \cite{Spielman2009}. Here we show that this momentum-space description can be extended to the case of unequal interactions, and under suitable conditions allows us to map the system into the chiral BF model at the quantum level Eq. \eqref{eq:cBF-H-enc-q} for experimental parameters realistic for $^{39}$K.

Specifically, we consider an atom with two internal states $\uparrow$ and $\downarrow$ separated by a frequency difference $\Delta$. Two Raman laser beams of wavelength $\lambda_R$ counter-propagating along the $x$ direction couple them, imparting a momentum along the coupling direction $k_x$ of $2k_R$, but leaving the momentum along the perpendicular directions ${\bf k_{\perp}}$ unaffected, see Fig. \ref{fig:Fig1}. Here $k_R=2\pi/\lambda_R$ is the single-photon Raman wavevector and $E_R=\hbar^2k_R^2/2 m$ is the corresponding recoil energy. In the following we work in dimensionless units and set $k_R$ and $E_R$ to $1$.

In the frame rotating at $\Delta$ and in the atomic basis $\ket{\sigma}$ formed by the states $\ket{\downarrow,k_x-1}$ and $\ket{\uparrow,k_x+1}$, the Hamiltonian of the system in the absence of interatomic interactions takes the form

\begin{equation}
\hat{H}_{\mathrm{kin}} = \int\frac{\mathrm{d}^3\mathbf{k}}{(2\pi)^3}\sum_{\sigma_1,\sigma_2}\hat{\varphi}_{\sigma_1}^{\dagger}(\mathbf{k}){\cal H}_{{\rm kin}, \sigma_1,\sigma_2}\hat{\varphi}_{\sigma_2}(\mathbf{k}),
\end{equation}
where the field operator $\hat{\varphi}_{\sigma}^{\dagger}(\mathbf{k})\,(\hat{\varphi}_{\sigma}(\mathbf{k}))$ creates (destroys) a particle in state $\ket{\sigma}$ and ${\cal H}_{{\rm kin},\sigma_1,\sigma_2}$ is an element of the $2\times2$ matrix
\par\noindent\begin{multline}
	{\cal H}_{\mathrm{kin}} =\cr\\
	\begin{pmatrix}
	{(k_x+1)^2 +\textbf{k}^2_{\perp}} - \delta_0/2 & \Omega/2 \\
	\Omega/2 & {(k_x-1)^2+\textbf{k}^2_{\perp}}  + \delta_0/2  \\
	\end{pmatrix},
	\label{eq:Hkin-matrix}
\end{multline}
where $\Omega$ and $\delta_0$ are the two-photon Rabi frequency and single-particle detuning of the Raman coupling beams written in dimensionless units.

For each value of $k_x$, ${\cal H}_{\mathrm{kin}}$ can be diagonalized by the unitary matrix
\begin{equation}
U_R(k_x) =
	\begin{pmatrix}
	\sin{\theta(k_x)} &  -\cos{\theta(k_x)}\\
	\cos{\theta(k_x)} &  \sin{\theta(k_x)} \\
	\end{pmatrix},
\label{eq:U}
\end{equation}
which relates the original $\ket{\sigma}$ basis to the Raman-dressed basis $\ket{\pm}$ \emph{via} a momentum-dependent basis transformation
\begin{equation}
\hat{\phi}_{\pm}(\mathbf{k})=\sum_{\sigma} U_{R\pm,\sigma}(k_x)\hat{\varphi}_\sigma(\bf{k})\label{eq:phi-dressedbasis},
\end{equation}
where $\hat{\phi}_{\pm}({\bf k})$ are the field operators in the dressed basis. In this basis, the eigenvalues read
\begin{equation}
E_{\pm}({\bf k})=\mathbf{k}_{\perp}^2+\mathcal{E}_{\pm}(k_x)=\mathbf{k}_{\perp}^2+k_x^2+1\pm\frac{1}{2}\tilde{\Omega}(k_x), \label{eq:E-dressedbasis}
\end{equation}
with $\tilde{\Omega}(k_x)=\sqrt{\Omega^2+\tilde{\delta}(k_x)^2}$ and $\tilde{\delta}(k_x)=\delta_0-4k_x$, and describe two energy bands of quasi-momentum ${\bf k}$. In Eq. \eqref{eq:U} the mixing angle $\theta(k_x)$ is momentum dependent and can be conveniently expressed as a function of the momentum-dependent polarization parameter $P=\tilde{\delta}/\tilde{\Omega}$. The latter describes the spin composition of the dressed states, with $\sin{\theta(k_x)}=\sqrt{(1-P)/2}$ and $\cos{\theta(k_x)}=\sqrt{(1+P)/2}$, and is an accessible experimental observable.

For $\Omega> 4$, the lowest energy band has a single minimum. Expanding $\mathcal{E}_-$ to second order in $k_x/\Omega$ around this minimum allows one to interpret the dressed-state atoms as synthetic particles of larger effective mass and subjected to a synthetic vector potential. For small detunings, it is proportional to $\delta_0$ \cite{Spielman2009}. Here, we perform instead such series expansion at a generic value of $k_x$. By rewriting $k_x=k_0+q$, we expand $\mathcal{E}_-$ in series of the small parameter $q/\tilde{\Omega}$ around $k_0$ and obtain

\begin{equation}
{\cal H}_{\mathrm{kin}}=\mathcal{E}_-(k_0)+\mathbf{k}^2_{\perp}+\frac{(q-A_{\mathrm{s}})^2}{m^*}+W_0+\mathcal{O}\left((q/\tilde{\Omega})^4\right)\label{eq:Hkin-q},
\end{equation}
where, along the $x$ direction, the dressed atoms experience a synthetic vector potential
\begin{equation}
A_{\mathrm{s}}=-m^*\left[k_0+\frac{\tilde{\delta}(k_0)}{\tilde{\Omega}(k_0)}-8\frac{\tilde{\delta}(k_0)\Omega^2}{\tilde{\Omega}(k_0)^5} q^2\right],\label{eq:A-q}
\end{equation}
have an effective mass
\begin{equation}
m^*=\left[1-4\frac{\Omega^2}{\tilde{\Omega}(k_0)^3}\right]^{-1},\label{eq:m*}
\end{equation}
and are subjected to an additional scalar potential $W_0=-A_{\mathrm{s}}^2/m^*$. To leading order, these expressions coincide with those obtained through the position-space approach. To second order in $q/\tilde{\Omega}$, they are analogous to the usual momentum-space results for $k_0=0$ \cite{Spielman2009}, except from the fact that $m^*$, $A_{\mathrm{s}}$ and $W_0$ now depend on the value of $k_0$. To third order in $q/\tilde{\Omega}$, $A_{\mathrm{s}}$ acquires a momentum dependence which can be interpreted as a $q$-dependent effective mass \cite{Khamehchi2017}.
Note that our series expansion remains accurate as long as $k_0$ is chosen equal or close to the center of mass momentum of the cloud. Indeed, the momentum spread of a weakly interacting BEC remains small compared to the Raman wavevector $k_R$ as long as the effective mass Eq. \eqref{eq:m*} is not too large.

The crucial ingredient to engineer the chiral BF model in its encoded form is to include interactions into this momentum-space picture. To this end, we start from the interaction Hamiltonian of the system in the $\ket{\sigma}$ basis
\begin{equation}
\hat{H}_{\mathrm{int}} = \int\frac{\mathrm{d}^3\mathbf{k}_1}{(2\pi)^3}\frac{\mathrm{d}^3\mathbf{k}_2}{(2\pi)^3}\frac{\mathrm{d}^3\mathbf{k}_3}{(2\pi)^3}\frac{\mathrm{d}^3\mathbf{k}_4}{(2\pi)^3}
\hat{\mathcal{V}}(\mathbf{k}_1,\mathbf{k}_2,\mathbf{k}_3,\mathbf{k}_4)\label{eq:Hint-def},
\end{equation}
where
\begin{multline}
\hat{\mathcal{V}}(\mathbf{k}_1,\mathbf{k}_2,\mathbf{k}_3,\mathbf{k}_4)=\frac{1}{2}\sum_{\sigma_1,\sigma_2}g_{\sigma_1,\sigma_2}\hat{\varphi}_{\sigma_1}^{\dagger}(\mathbf{k}_4)\hat{\varphi}_{\sigma_2}^{\dagger}(\mathbf{k}_3)\\
\hat{\varphi}_{\sigma_1}(\mathbf{k}_2)\hat{\varphi}_{\sigma_2}(\mathbf{k}_1)\delta^3(\mathbf{k}_4+\mathbf{k}_3-\mathbf{k}_2-\mathbf{k}_1)\label{eq:Hint-Vintdef}.
\end{multline}

Eqs. \eqref{eq:Hint-def} and \eqref{eq:Hint-Vintdef} describe collisions between atoms with incoming momenta ${\bf k_1}$, ${\bf k_2}$, and outgoing momenta ${\bf k_3}$, ${\bf k_4}$, where the delta function ensures momentum conservation. In this expression, $g_{\uparrow\uparrow}$, $g_{\uparrow\downarrow}$, and $g_{\downarrow\downarrow}$ are the coupling constants describing the interactions between the two atomic internal states in the absence of Raman coupling.

Inspired by Refs. \cite{Search2001, Williams2012} and by our recent theoretical and experimental work on coherently-coupled BECs \cite{Cabedo2021a, Cabedo2021b, Sanz2022}, as a next step we express $\hat{\mathcal{V}}$ in the dressed basis $\ket{\pm}$ and restrict ourselves to the lowest energy band. This approximation is valid when the energy gap to the $\ket{+}$ band is much larger than all interaction energies of the problem, so that band-coupling processes are negligible. In these conditions, the effective interaction Hamiltonian in the lowest energy band includes
\par\noindent\begin{multline}
 \hat{\mathcal{V}}_{\mathrm{eff}}(\mathbf{k}_1,\mathbf{k}_2,\mathbf{k}_3,\mathbf{k}_4)=\frac{1}{2}\tilde{g}_{\mathrm{eff}}({\bf k_1},{\bf k_2},{\bf k_3},{\bf k_4})\\
 \hat{\phi}^\dagger(\mathbf{k}_4)\hat{\phi}^\dagger(\mathbf{k}_3)
 \hat{\phi}(\mathbf{k}_2)\hat{\phi}(\mathbf{k}_1)\delta^3(\mathbf{k}_4+\mathbf{k}_3-\mathbf{k}_2-\mathbf{k}_1).
\label{eq:Veff}
\end{multline}
Here, we have introduced the simplified notation $\hat{\phi}\equiv\hat{\phi}_-$ and defined $\tilde{g}_{\mathrm{eff}}$, the effective coupling constant describing the interactions between the Raman-dressed atoms. This coupling constant is momentum dependent, and depends only on the projection of the momenta of the colliding atoms along the $x$ axis, $k_{i,x}$. It reads
\par\noindent\begin{multline}
	\tilde{g}_{\mathrm{eff}}(k_{1,x},k_{2,x},k_{3,x},k_{4,x})=\sum_{\sigma_1,\sigma_2}g_{\sigma_1,\sigma_2}\\
U_{R-,\sigma_1}(k_{4,x})U_{R-,\sigma_1}^\dagger(k_{2,x})U_{R-,\sigma_2}(k_{3,x})U^\dagger_{R,-,\sigma_2}(k_{1,x}).
\label{eq:gefftilde}
\end{multline}
Thus, the coupling leads to an effective modification of the scattering properties of the dressed states, an effect previously explored only in the cases $g_{\uparrow\uparrow}\neq g_{\downarrow\downarrow}$, $k_R=0$ \cite{Search2001, Nicklas2011, Sanz2022}, and $g_{\uparrow\uparrow}=g_{\downarrow\downarrow}= g_{\uparrow\downarrow}$ and $k_R\neq 0$ \cite{Williams2012}.

As with ${\cal H_{\mathrm{kin}}}$, $\tilde{g}_{\mathrm{eff}}$ can be expanded in series around any momentum $k_0$. Writing the momenta of the atoms along $x$ as $k_{i,x}=k_0+q_{i,x}$ and expanding in the small parameter $q_{i,x}/\tilde{\Omega}$, we obtain to first order
\par\noindent\begin{align}
&\tilde{g}_{\mathrm{eff}}(k_{1,x},k_{2,x},k_{3,x},k_{4,x})=\label{eq:geff}\\
&\tilde{g}_{\mathrm{eff}}(k_0,k_0,k_0,k_0)+\lambda\frac{\tilde{\Omega}(k_0)}{2 m^*}\sum_{i=1}^4\frac{q_{i,x}}{\tilde{\Omega}(k_0)}+{\cal O}\left((q_i/\tilde{\Omega})^2\right)\nonumber.
\end{align}
The zeroth order expansion coefficient is
\par\noindent\begin{multline}
\tilde{g}_{\mathrm{eff}}(k_0,k_0,k_0,k_0)=g_{\uparrow\uparrow}\cos^4{\theta(k_0)}\\
+g_{\downarrow\downarrow}\sin^4{\theta(k_0)}+\frac{1}{2}g_{\uparrow\downarrow}\sin^2{2\theta(k_0)}=g_{\text{eff}}(P_0),
\end{multline}
where $P_0=P(k_0)$ is the polarization parameter for $k_0$ and $g_{\text{eff}}(P)=[g_{\uparrow\uparrow}(1 + P)^2 + g_{\downarrow\downarrow}(1-P)^2 + 2g_{\uparrow\downarrow}(1-P^2)]/4$ is the effective interaction strength commonly used in the $k_R=0$ case \cite{Search2001, Sanz2022, Nicklas2011}. The first order expansion coefficient is
\begin{eqnarray*}
\lambda&=&\frac{4m^*}{\tilde{\Omega}(k_0)}\left[\frac{\tilde{\delta}(k_0)}{\tilde{\Omega}(k_0)}g_{\text{eff}}(P_0) + g_{\downarrow\downarrow}\sin^4{\theta(k_0)}-g_{\uparrow\uparrow}\cos^4{\theta(k_0)}\right]\nonumber\\
&=&\frac{m^*}{\tilde{\Omega}(k_0)}\left[\frac{4\tilde{\delta}(k_0)}{\tilde{\Omega}(k_0)}g_{\text{eff}}(P_0) + g_{\downarrow\downarrow}(1-P_0)^2-g_{\uparrow\uparrow}(1+P_0)^2\right].\nonumber
\end{eqnarray*}

Combining Eqs. \eqref{eq:Veff} and \eqref{eq:geff}, and transforming back to position space, the interaction Hamiltonian truncated to the lowest energy band can be rewritten as $\hat{H}_{\mathrm{int}}=\int\mathrm{d}^3\mathbf{r}\hat{\cal{H}}_{\mathrm{int}}$, with
\begin{widetext}
    \begin{align}
    \hat{\cal{H}}_{\mathrm{int}}
    & \approx \frac{1}{2}\int\frac{\mathrm{d}^3\mathbf{q}_1}{(2\pi)^3}\frac{\mathrm{d}^3\mathbf{q}_2}{(2\pi)^3}\frac{\mathrm{d}^3\mathbf{q}_3}{(2\pi)^3}\frac{\mathrm{d}^3\mathbf{q}_4}{(2\pi)^3}
    \left[g_{\mathrm{eff}}(P_0)+\lambda\frac{\tilde{\Omega}(k_0)}{2 m^*}\sum_{i=1}^4\frac{\mathbf{q}_i\cdot\mathbf{e}_1}{\tilde{\Omega}(k_0)}\right]
    \hat{\phi}^{\dagger}(\mathbf{q}_3)\hat{\phi}^{\dagger}(\mathbf{q}_4)\hat{\phi}(\mathbf{q}_1)\hat{\phi}(\mathbf{q}_2)\label{eq:Hintfinal}\\
    & \exp\left[i\left(\mathbf{q}_1+\mathbf{q}_2-\mathbf{q}_3-\mathbf{q}_4\right)\cdot\mathbf{r}\right]= \frac{1}{2}g_{\mathrm{eff}}(P_0)\hat{\phi}^{\dagger}(\mathbf{r})\hat{n}(\mathbf{r})\hat{\phi}(\mathbf{r})
		+\frac{\lambda}{2im^*}\hat{\phi}^{\dagger}(\mathbf{r})
    \left[\hat{\phi}^{\dagger}(\mathbf{r})\partial_x\hat{\phi}(\mathbf{r})- [\partial_x\hat{\phi}^\dagger(\mathbf{r})]\hat{\phi}(\mathbf{r})\right]\hat{\phi}(\mathbf{r}).\nonumber
    \end{align}
\end{widetext}

This interaction Hamiltonian density includes, besides the usual density-density interaction term, an additional contribution proportional to $\lambda$. Recognizing the (dimensionless) normal-ordered current operator in the absence of external gauge fields,
\begin{equation}
\hat{J}(\mathbf{r})=\frac{1}{i m^*}\left[\hat{\phi}^\dagger(\mathbf{r})\partial_x\hat{\phi}(\mathbf{r})- [\partial_x\hat{\phi}^\dagger(\mathbf{r})]\hat{\phi}(\mathbf{r})\right],
\end{equation}
we see that this contribution corresponds to $\cal{H}_{\mathrm{int}}^{\mathrm{chiral}}$. Thus, to first order in $\lambda$, Eq. \eqref{eq:Hintfinal} is identical to the quantum version of the interaction term of the chiral BF theory after encoding Eq. \eqref{eq:cBF-H-enc-int}
\begin{equation}
{\cal \hat{H}}_{\mathrm{int}} \approx \frac{1}{2}g_{\text{eff}}(P_0):\hat{n}^2:+\frac{\lambda}{2} :\hat{J}\hat{n}:+\mathcal{O}(\lambda^2),
\label{eq:cBF-H-enc-qR}
\end{equation}
which to $\mathcal{O}(\lambda^2)$ is equivalent to the Kundu linear anyon model \footnote{Note that the interaction Hamiltonian density of Eq. \eqref{eq:cBF-H-enc-qR} is equivalent to the second-quantized bosonic Hamiltonian Eq. (2) of Ref. \cite{Kundu1999} for $c= g_{\mathrm{eff}} (P_0)m^*/2$ and $\kappa=-\lambda/2$.}.

To derive the complete Hamiltonian of the encoded chiral BF theory, we need to reproduce as well its kinetic term. Starting from the expression of $\hat{\cal{H}}_{\mathrm{kin}}$ Eq. \eqref{eq:Hkin-q}, we achieve this by setting $\tilde{\delta}=0$. This choice corresponds to $k_0=\delta_0/4$ and implies a balanced superposition of the states $\uparrow$ and $\downarrow$ ($P=0$). It cancels the $q$-dependence of the external vector potential $A_s$, which can then be gauged away. Dropping as well constant energy terms from Eq. \eqref{eq:Hkin-q} and combining the result with the expression of $\hat{\cal{H}}_{\mathrm{int}}$ Eq. \eqref{eq:Hintfinal}, we obtain the effective Hamiltonian
\par\noindent\begin{multline}
\hat{H}_{\mathrm{eff}}
\approx
\int \textrm d^3 \mathbf r\hat{\phi}^\dagger(\mathbf{r})\Bigg[-\nabla_{\perp}^2-\frac{\partial_x^2}{m^*}+\\
+\frac{1}{2}\left(g_1-\frac{\lambda\delta_0}{2} \right)\hat{n}(\mathbf{r})
+ \frac{\lambda}{2} \hat{J}(\mathbf{r})\Bigg]\hat{\phi}(\mathbf{r}),
\label{eq:gpe-H-Raman}
\end{multline}
where $g_1=g_{\text{eff}}(0)$. Around any momentum $k_0$, Eq. \eqref{eq:gpe-H-Raman} describes a system of bosonic particles of effective mass $m^*$ along the $x$ direction that have not only standard density-density interactions of coupling strength $g_1-\lambda\delta_0/2$, but also the characteristic current-density interactions of the chiral BF theory. Indeed, this expression is identical to the encoded version of the chiral BF Hamiltonian Eq. \eqref{eq:cBF-H-enc-q} for $\hat{\tilde{V}}(\hat n)=(g_1-\lambda\delta_0/2)\hat{n}^2$.

In the weakly interacting regime $\Omega\gg g_{\uparrow\uparrow}n,\, g_{\uparrow\downarrow} n$ and $g_{\downarrow\downarrow} n$ considered in this work, Eq. \eqref{eq:gpe-H-Raman} can also be rewritten as
\par\noindent\begin{multline}
\hat{H}_{\text{eff}}=\int{\textrm d^3 \mathbf r\hat{\phi}^\dagger(\mathbf{r})\bigg[-\nabla_{\perp}^2-\frac{1}{m^*}\left(\partial_x+i\frac{\lambda} {2}\hat{n}(\mathbf{r})\right)^2}\\
+\frac{1}{2}\left(g_1-\frac{\lambda\delta_0}{2}\right)\hat{n}(\mathbf{r})\bigg]\hat{\phi}(\mathbf{r})+\mathcal{O}(\lambda^2),
	\label{eq:cBF-H-Raman}
\end{multline}
where we have added a (negligible) three-body interaction term which, in dimensionless units, reads ${\hat{\mathcal{H}}^{3B}_{\text{int}}=\lambda^2:\hat{n}^3:/(4m^*)}$. Equation \eqref{eq:cBF-H-Raman} is equivalent to the quantum version of the encoded chiral BF Hamiltonian Eq. \eqref{eq:cBF-H-enc-An} and describes a bosonic field minimally coupled to the density-dependent vector potential $\hat{{\cal A}}=-\lambda \hat{\phi}^{\dagger}\hat{\phi}/2=-\lambda\hat{n}/2$. Thus, this result connects our momentum-space derivation of the effective Hamiltonian to the position-space derivation of Ref. \cite{Edmonds2013}, and shows that the latter is the classical field limit of the former.

In conclusion, in this Section we have shown that for $\tilde{\delta}=0$ $(P=0)$ the lowest dressed state of a Raman-coupled BEC with unequal interactions effectively realizes the encoded version of the chiral BF theory introduced in Section \ref{Sec3}. We have obtained this result by microscopically deriving the effective interaction Hamiltonian of the system restricted to the lowest energy band. Thus, in contrast to the original position-space treatment of Ref. \cite{Edmonds2013}, which focuses on the connection to a density-dependent vector potential, our approach highlights the microscopic origin of the chiral (current-density) interaction term. Specifically, we have shown that this term directly reflects the momentum-dependence of the interactions between Raman-dressed states when $g_{\uparrow\uparrow}\neq g_{\downarrow\downarrow}$, and results in a system where $\cal{H}_{\mathrm{int}}$ breaks Galilean invariance already at the lowest order in momentum.

Our method allows us to extend the validity of the mapping to the encoded chiral BF theory from the classical to the quantum regime, and remains valid beyond the mean-field approximation. Moreover, we show that the mapping extends to moderate values of the Raman coupling strength, well beyond the regime of applicability of the position-space approach, provided the relation $k_0=\delta_0/4$ (i.e. $\tilde{\delta}=0$ and $P=0$) is fulfilled. This point was key for our experimental realization of the model \cite{Froelian2022}. In the next Section, we investigate more in detail the anomalous properties of chiral Bose-Einstein condensates, i.e. BECs described by the effective Hamiltonian of Eq. \eqref{eq:cBF-H-Raman}, and relate them to the phenomenology of the underlying one-dimensional topological gauge theory.

\section{Properties of a chiral BEC from a gauge theory perspective\label{Sec5}}
Chiral BECs have been predicted to display a number of unconventional properties, including the existence of chiral solitons \cite{Edmonds2013, Dingwall2018, Dingwall2019, Bhat2021}, density-dependent persistent currents and anomalous expansion dynamics \cite{Edmonds2013}, unconventional collective modes \cite{Edmonds2015, Zheng2015, Chen2021}, and peculiar vortex patterns \cite{Butera2016, Edmonds2020}. Previous works viewed these properties as a result of an anomalous current term in the equations of motion originating from the density-dependent vector potential $\cal A$ of the encoded Hamiltonian. Here, we show that the correspondence to the underlying chiral BF gauge theory provides a more intuitive interpretation of these results. Specifically, we review the main observables of the chiral BF theory that can be accessed by performing measurements on a chiral BEC, and discuss their potential experimental observation.

\emph{Chiral solitons.} One of the defining properties of the chiral BF theory is the existence of chiral soliton solutions for the matter field \cite{Aglietti1996, Jackiw1997, Griguolo1998}. These are self-bound wavepackets of the matter field that propagate without dispersion only when moving along one direction. On the contrary, wavepackets with the opposite center of mass momentum cannot form solitons and spread. Thus, chiral solitons are collective excitations of the one-dimensional system and can be seen as “particles” which only exist for a given propagation direction.

Solitonic solutions of the matter field can be dark solitons, i.e. dips in the density profile of the matter field that are accompanied by an abrupt change of its phase, or bright solitons, i.e. self-bound matter wavepackets that propagate without dispersion. Both are solutions of the classical equation of motion for the matter field of the encoded BF Hamiltonian Eq. \eqref{eq:cBF-H-enc}, which is an effective one-dimensional extended Gross-Pitaevskii equation (eGPE) of the form \cite{Aglietti1996, Edmonds2013}
\begin{equation}
i\dot{\phi}=-\left(\partial_x+i\frac{\lambda}{2} n\right)^2\phi+\left(g n+\frac{\lambda}{2} J\right) \phi+\mathcal{O}(\lambda^2).\label{eq:cBF-eGPE}
\end{equation}
Here, we use the same units as in Section \ref{Sec4}. By performing a Jordan-Wigner transformation $\Phi=\exp(i \int_{-\infty}^x \mathrm{d}\xi\,{\lambda n/2})\phi$ that eliminates the density dependence from the kinetic term, we obtain
\begin{equation}
i\dot{\Phi}=-\Phi''+(g n+\lambda J) \Phi+\mathcal{O}(\lambda^2)\label{eq:cBF-eGPE-2}.
\end{equation}
If we consider a matter wavepacket of center of mass momentum $k$, its group velocity in dimensionless units is $v=2k$ and the current reads $J=2kn=n v$. Then, Eq. \eqref{eq:cBF-eGPE-2} becomes completely analogous to the usual Gross-Pitaevskii equation but with $\tilde{g}=g+2\lambda k$. The additional momentum-dependent part of the coupling constant corresponds to the chiral interaction term of the matter field in the encoded theory $\mathcal{H_{\mathrm{int}}}^{\mathrm{chiral}}=\lambda kn^2$, see Section \ref{Sec3}. When $\tilde{g}<0$, the system supports a bright soliton solution: a self-bound matter wavepacket that propagates without dispersion because the attractive non-linearity compensates for the dispersion caused by quantum pressure. If the condensate at rest has $g\geq0$, reversing its group velocity will yield a repulsively interacting system ($\tilde{g}>0$) where no soliton solutions exist \cite{Edmonds2013}, as we have indeed observed experimentally \cite{Froelian2022}. Thus, the bright soliton solution of Eq. \eqref{eq:cBF-eGPE-2} is chiral, and constitutes the many-body analogue of edge states in quantum Hall systems. It directly reflects the chirality of the underlying gauge theory, which can be traced back to the self-dual term $\dot{\mathcal{B}}\mathcal{B}'$ of the chiral BF model before encoding, Eq. \eqref{eq:cBF}.

\emph{Electric field.} In its encoded form, the chiral BF theory involves only matter fields. However, the relation between the gauge and matter classical fields derived in Sections \ref{Sec2} and \ref{Sec3} translates into a relation between the expectation values of the corresponding quantum fields. Specifically, one has $\langle \hat{E}\rangle =\lambda \langle\dot{\hat{n}}\rangle$ and can therefore determine the chiral BF electric field by measuring the temporal evolution of the matter density. Changes in the density induce an electric force that acts back into the chiral BEC and endows it with rich dynamics, especially when the system is confined in a harmonic trap.

For instance, Refs. \cite{Edmonds2015, Zheng2015, Chen2021} showed that the current-density term of the eGPE Eq. \eqref{eq:cBF-eGPE} leads to a coupling of the monopole and dipole collective modes of the system, eventually leading to chaotic dynamics. Here, we interpret this coupling in terms of the induced chiral BF electric field. In a chiral BEC, the compression of the gas due to the excitation of the breathing mode generates a synthetic electric field. The corresponding electric force displaces the center of mass of the cloud and excites the dipole mode. Thus, the coupling of the monopole and dipole modes is a direct manifestation of the back action between matter and gauge fields.

An even stronger manifestation of the BF electric field is the distortion of the density profile of the cloud upon expansion of the chiral BEC into an optical waveguide, an effect that we have recently observed experimentally \cite{Froelian2022} following the theoretical prediction of Ref. \cite{Edmonds2013}. In this situation, the density of the gas drops in the center of the cloud and increases on its edges, leading to an inhomogeneous distribution of electric forces. These forces skew the density profile of the chiral BEC, which develops an asymmetric shape during the expansion. Within the chiral BF theory, the only source of asymmetry of the density distribution is the chiral BF electric field. Other sources of asymmetry, stemming from higher order momentum corrections of the kinetic term \cite{Khamehchi2017} go beyond the mapping of the system into the encoded chiral BF Hamiltonian, see Section \ref{Sec4}. Thus, tracing the different sources of asymmetry during the expansion of the system is a powerful means of benchmarking the validity of the mapping, as we discuss in detail in the next Section.

\section{Numerical validation for realistic experimental conditions\label{Sec6}}

\begin{figure*}[t]
\includegraphics{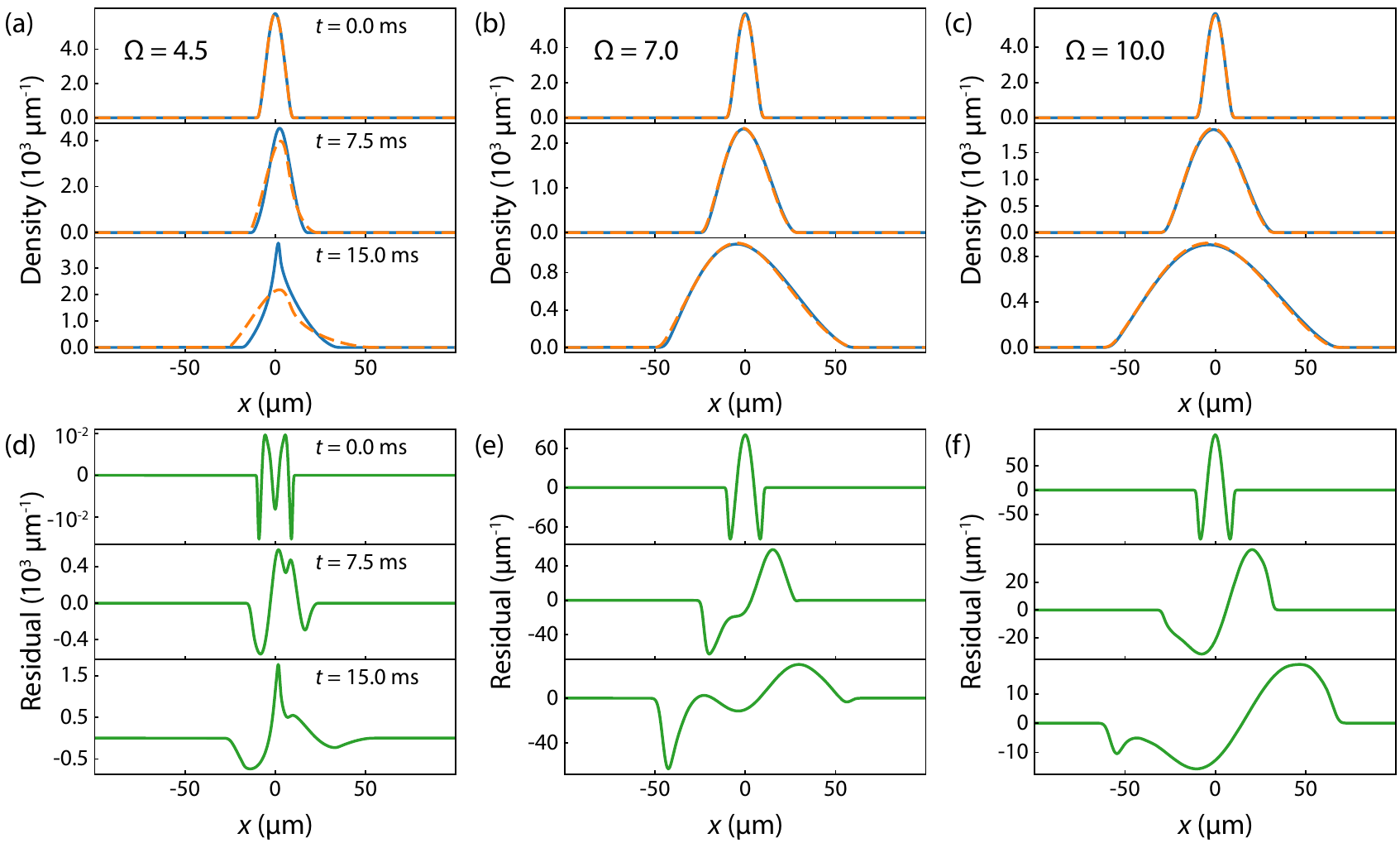}
\caption{\label{fig:Fig2} Numerical comparison of the expansion dynamics for the effective single-component model corresponding to the encoded chiral BF theory and the complete two-component Raman-coupled system. (a)-(c) Integrated density profiles of the effective model (blue solid lines) and of the full system (orange dashed lines) for (a) $\Omega=4.5$, (b) $\Omega=7.0$, and (c) $\Omega=10.0$ after $0$, $7.5$, and $15$ ms of expansion in the optical waveguide. The dynamics of both are qualitatively similar for all values of $\Omega$. For $\Omega\geq7.0$, the two density profiles are practically indistinguishable as shown by the difference in the profiles given in (d)-(f).}
\end{figure*}

In the previous Sections, we have derived a Hamiltonian encoding of the chiral BF theory that includes matter-only degrees of freedom and is amenable to quantum simulation with ultracold quantum gases (Section \ref{Sec3}), and we have shown how to implement it in Raman-coupled BECs with imbalanced intrastate interactions (Section \ref{Sec4}). Moreover, we have discussed how to reveal the key phenomenology of the gauge theory in a chiral BEC using realistic experimental observables (Section \ref{Sec5}). Specifically, we have seen that letting the condensate expand in an optical waveguide results in an asymmetric density profile which reveals the electric field of the chiral BF theory before encoding. In this Section, we use this characteristic feature to verify the quality of the mapping of the Raman-coupled BEC into Eq. \eqref{eq:cBF-H-enc} as a function of the relevant experimental parameters.

We consider the experimental conditions of Ref. \cite{Froelian2022}, i.e. a chiral $^{39}$K BEC composed by two $m_F$ magnetic sublevels of the $F=1$ hyperfine manifold $\ket{\uparrow}\equiv\ket{F=1,m_F=0}$ and $\ket{\downarrow}\equiv\ket{F=1,m_F=1}$. At a magnetic field $B_0=397.01$ G, their interactions are parameterized by the scattering lengths $a_{\uparrow\uparrow}/a_0=1.3$, $a_{\downarrow\downarrow}/a_0=252.7$, and $a_{\uparrow\downarrow}/a_0=-6.3$ \cite{Roy2013}, where $a_0$ is the Bohr radius, and that in the dimensionless units introduced in Section \ref{Sec4} are related to the corresponding coupling constants by $g_{\sigma_1\sigma_2}=8\pi a_{\sigma_1\sigma_2}$. The two Raman coupling beams are counter-propagating and their wavelength $\lambda_R=768.97$~nm corresponds to the potassium tune-out value, for which the scalar light shift cancels and no trapping potential is produced. To confine the atoms, a harmonic spin-independent trapping potential of frequencies ($\omega_x, \omega_y ,\omega_z)/2\pi=(70, 147, 99)$ Hz is included.

\begin{figure}[t]
\includegraphics{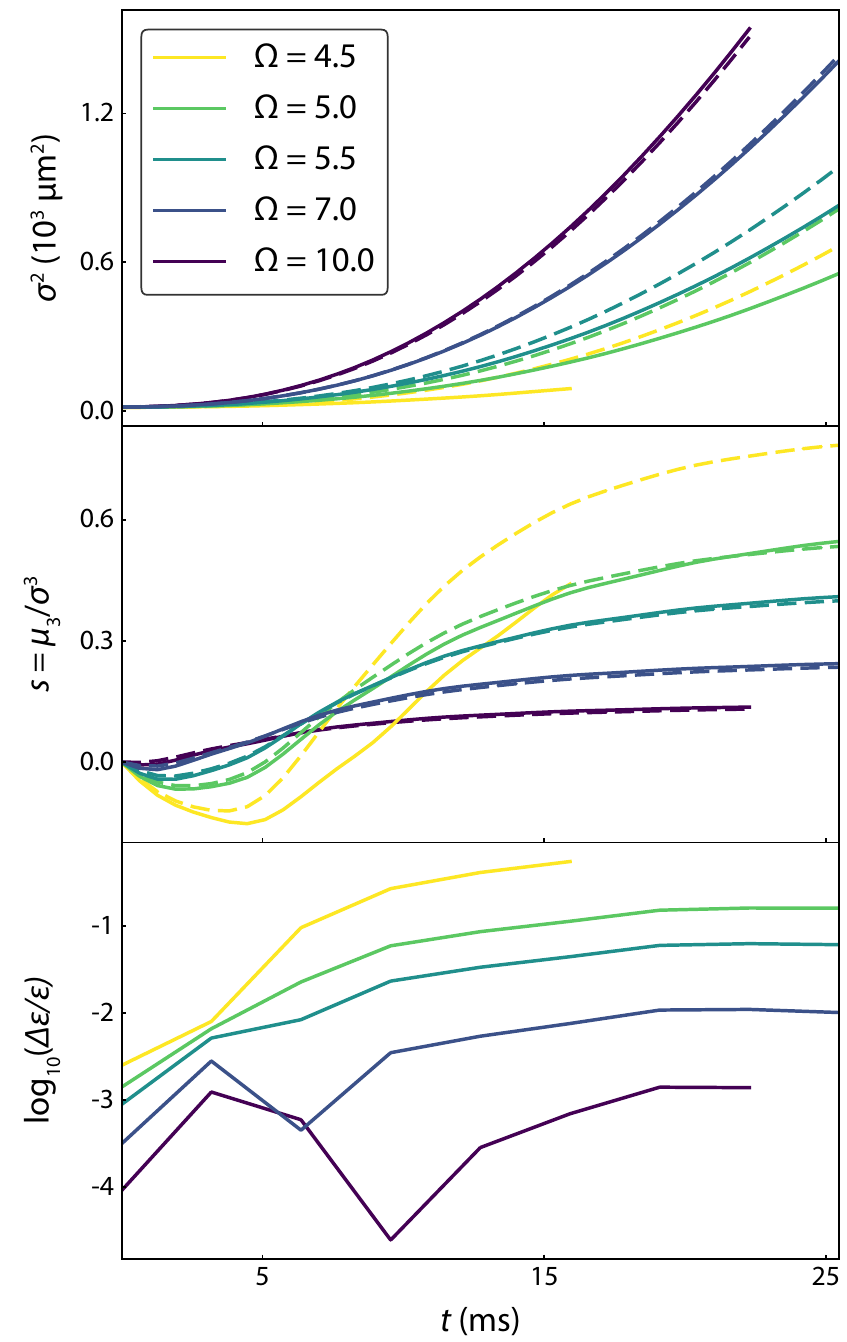}
\caption{\label{fig:Fig3} Validity of the mapping to the encoded chiral BF Hamiltonian. Top and middle panels: variance $\sigma^2$ and skewness $s=\mu_3/\sigma^3$ vs. expansion time $t$ for the effective single-component system (dashed lines) and the full two-component system (solid lines), for different values of $\Omega$. For $\Omega\geq 7.0$ the two models become practically indistinguishable. Bottom panel: relative error of the next order energy corrections $\Delta \cal{E}/\cal{E}$, which remains below $1$~\% for $\Omega\geq7.0$.}
\end{figure}

We aim at determining the experimental parameters for which the dynamics of this system are faithfully described by the encoded chiral BF theory, i.e. by the effective single-component Hamiltonian of Eq. \eqref{eq:gpe-H-Raman} and the corresponding extended Gross-Pitaevskii equation \footnote{We therefore do not include the three-body term $\hat{\cal{H}}_{\mathrm{chiral}}^{\mathrm{3B}}$, which contributes less than $10^{-5}$ as a fraction of the interaction energy for all parameters considered here.}. To this end, we compare the predictions of the effective single-component model to those of the complete two-component system for different values of the Rabi coupling $\Omega$, restricting ourselves to the regime where the dispersion relation has a single minimum ($\Omega>4$). Specifically, we compute the ground state of the system in the trap using imaginary time evolution. We set $\delta_0=0$ which gives an initial center of mass momentum of the BEC approximately equal to $k_0=0$ \footnote{In the ground state, the BEC has no mechanical momentum, i.e. its center of mass momentum corresponds to that of the minimum of the dispersion relation. This value is slightly shifted from the single particle prediction ($k_x=0$ for $\delta_0=0$) due to the effect of the differential mean-field energy shift on the Raman detuning.}. We numerically simulate the time-dependent dynamics of the cloud upon release from the harmonic trap into an optical waveguide that tightly confines the atoms in the transverse directions. During the process, the Raman beams are kept on.  As discussed in Section \ref{Sec5}, we expect the atomic density distribution to become asymmetric during the expansion, revealing the effect of the electric field of the chiral BF theory. We characterize this asymmetry \emph{via} the  second and third central moments of the distribution, $\sigma^2$ (the variance) and $\mu_3$, from which we can quantify the skewness of the distribution $s=\mu_3/\sigma^3$. In the experiment of Ref. \cite{Froelian2022}, the optical waveguide is formed by a single optical dipole trap beam of transverse trapping frequencies $\left(\omega_y, \omega_z\right)/2\pi=\left(129, 99\right)$~Hz, where the asymmetry between the two directions is due to the effect of gravity. A weak axial potential $\omega_x/2\pi=5$ Hz stemming from the residual curvature of the magnetic field is added to it. In all simulations, we consider exactly this combined optical/magnetic potential to facilitate the comparison with the experimental data. Since the trapping potential does not have cylindrical symmetry, we perform full three-dimensional simulations. We exploit the package XMDS2~\cite{Dennis2013}, and employ an adaptive time step Runge-Kutta $4(5)$ algorithm using $4096$ grid points with $0.058$ $\mu$m spacing along the $x$-axis and $32$ grid points with $0.675$ $\mu$m spacing along each transverse axis. For the results presented here, we integrate out the transverse axes.

\begin{figure}[t]
\includegraphics{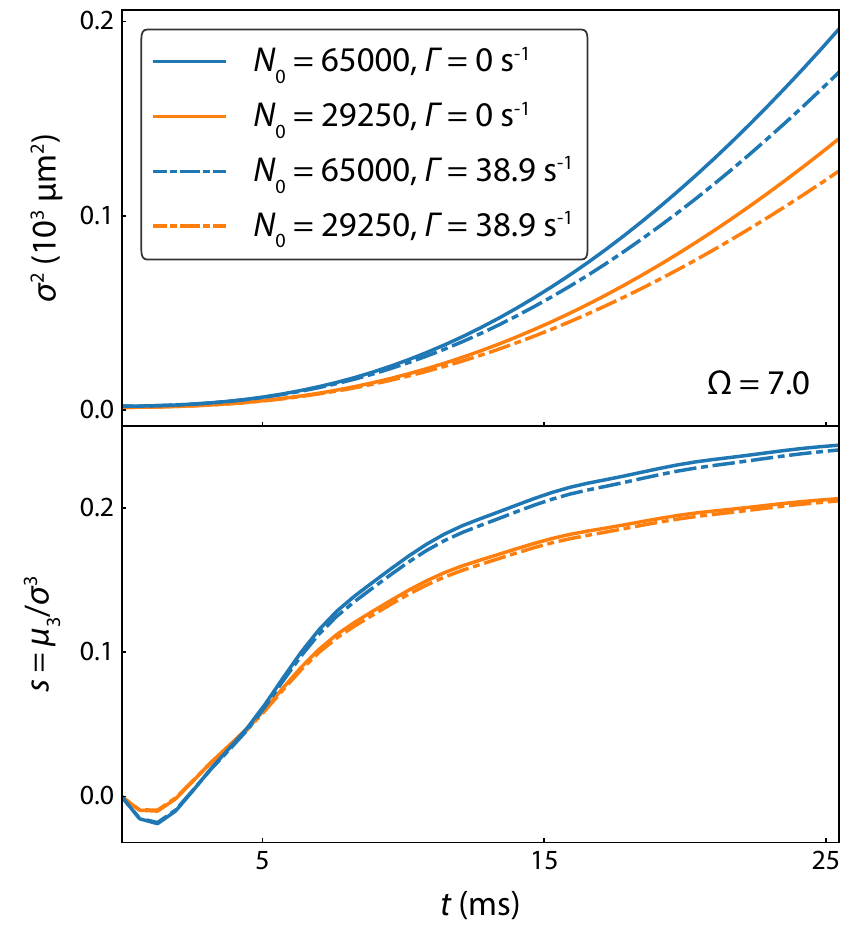}
\caption{\label{fig:Fig4} Effect of atomic losses on the asymmetric expansion of a chiral BEC. Variance $\sigma^2$ (top panel) and skewness $s=\mu_3/\sigma^3$ (bottom panel) of the effective single-component model in the presence and absence of an atom loss rate $\Gamma$, for $\Omega=7.0$ and two different values of the initial atom number $N_0$. The simulations show the robustness of the skewness against atom loss.}
\end{figure}

In a first series of simulations, we fix a total atom number $N=65000$. Figure \ref{fig:Fig2} (a)-(c) shows the evolution of the density profiles of the effective single-component system (corresponding to the encoded chiral BF theory) and of the full two-component Raman-coupled system for three values of $\Omega$. Moreover, Figs. \ref{fig:Fig2} (d)-(f) show the difference between the single- and two-component density profiles. We see that for all considered values of $\Omega$, the effective chiral BF model qualitatively reproduces the behavior of the full Raman-coupled system. Remarkably, this is even true for values of $\Omega$ barely above $4$, i.e. just at the entrance of the single-minimum regime. In Fig.~\ref{fig:Fig3} we compute $\sigma^2$ and the skewness of the density distributions over time for five values of $\Omega$ (top and middle panels). For $\Omega\geq7$, the predictions of the effective single-component model become practically indistinguishable from those of the complete system for all times, showing that the Raman-coupled BEC faithfully implements the encoded chiral BF theory in this regime. To quantify the quality of the implementation, we compute the next order corrections in energy for the effective model as a fraction of the total energy (discounting the transverse kinetic energy and trap contributions) $\Delta \cal{E} /\cal{E}$ (bottom panel). For $\Omega\geq7$, we get an energy discrepancy of $<1$~\% for all computed times. For comparison, in the long time limit the position-space approach of Ref. \cite{Edmonds2013} yields a fractional error with respect to Eq. \eqref{eq:cBF-H-Raman} equal to the effective mass $m^*$, which exceeds $100$~\% in this regime. This confirms that our microscopic momentum-space picture is essential to extend the validity of the mapping of the Raman-coupled system into the chiral BF theory to a regime that can be accessed in current experiments.

In a second series of simulations, we investigate the effect of atomic losses induced by the Raman-coupling beams on the expansion of the cloud. Since they remain experimentally unavoidable even at moderate values of the Rabi frequency \cite{Froelian2022}, our goal is to identify an experimental observable for the cloud asymmetry that is robust against them. To this end, we perform simulations of the effective single-component model in the presence and absence of the a phenomenological loss rate term $i\Gamma/2$. We set $\Gamma/\Omega=10^{-4}$, corresponding to the value predicted theoretically \cite{Wei2013} and measured experimentally in Ref. \cite{Froelian2022}. We consider $\Omega=7.0$ and two different values of the initial atom number $N_0$. As depicted in Fig. \ref{fig:Fig4}, both the variance and the skewness are affected by changes in $N_0$. In contrast, while the variance is affected by atomic losses, the skewness parameter is robust against them, and is thus a good experimental observable to quantify the asymmetry of the density distribution.

\begin{figure}
\includegraphics{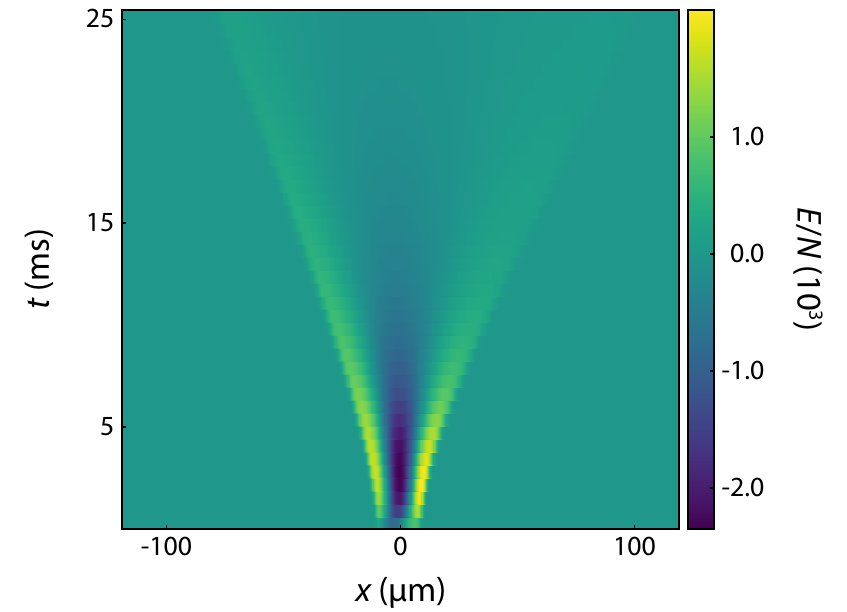}
\caption{\label{fig:Fig5} The chiral BF electric field generated by the time variation of the atomic density during the expansion for $\Omega=7.0$ and $N=65000$.}
\end{figure}

Finally, we evaluate the chiral BF electric field generated by the time variation of the atomic density during the expansion. As we discussed in Sections \ref{Sec3} and \ref{Sec5}, although the chiral BF gauge fields are eliminated in the encoded formulation of the chiral BF theory, the relation between matter and gauge fields imposed by the local symmetry constraint $\langle\hat{E}\rangle=\lambda \langle\dot{\hat{n}}\rangle$ allows us to recover the value of the chiral BF electric field, which we show in Fig. \ref{fig:Fig5} for $\Omega=7.0$ and $N=65000$.

In conclusion, we have shown that the encoded chiral BF theory can be realistically implemented in a Raman-coupled $^{39}$K BEC with imbalanced intrastate interactions, as investigated in Ref. \cite{Froelian2022}. We have quantified the validity of the mapping by simulating the expansion dynamics of the system in an optical waveguide, comparing the behavior predicted by an effective single-component model corresponding to the encoded chiral BF Hamiltonian to that of the complete two-component Raman-coupled system. We have shown that the mapping extends to moderate values of the Rabi frequency well beyond the regime of validity of the position-space approach and can be readily accessed in current experiments. We have found that the asymmetry of the density distribution, which reveals the chiral BF electric field of the theory before encoding, can be robustly characterized by its skewness parameter and is resilient to relevant experimental perturbations such as atom losses. Our numerical results provide a detailed theoretical support for the experiments of Ref. \cite{Froelian2022}. Moreover, a similar strategy could be exploited to benchmark the quality of the implementation of other encoded theories, such as the chiral BF theory with closed boundary conditions or the Chern-Simons theory in two dimensions \cite{Valenti2020}.

\section{Conclusions and outlook\label{Sec7}}

In this work, we have described and analyzed a scheme for the quantum simulation of the chiral BF theory, a topological gauge theory resulting from the dimensional reduction of the U(1) Abelian Chern-Simons gauge theory from two to one spatial dimensions, that we have recently demonstrated experimentally \cite{Froelian2022}. Our scheme is based on encoding the gauge degrees of freedom into a matter field with unconventional interactions using the local conservation laws of the theory, and thus preserves the gauge invariance of the dynamics by construction. It is equivalent to the strategy used to implement the Schwinger lattice gauge theory in trapped-ion digital quantum simulators \cite{Martinez2016, Muschik2017, Kokail2019}, but applied here to the realization of a continuum topological gauge theory in a neutral-atom analog quantum simulator. Using the so-called Faddeev-Jackiw first-order approach, we have shown that interactions in the encoded Hamiltonian are chiral, i.e. linearly dependent on the center of mass of the matter particles \cite{Rabello1996, Aglietti1996}. Then, building upon the seminal proposal of Edmonds \emph{et al.} \cite{Edmonds2013}, we have demonstrated that they can be engineered in two-component Raman-coupled Bose-Einstein condensates with imbalanced interactions for realistic experimental parameters, and that the mapping of this system to the chiral BF theory extends to the quantum regime. Finally, we have discussed how to extract the chiral BF gauge fields by performing measurements on the encoded system, and benchmarked the validity of our scheme for the experimental parameters of Ref. \cite{Froelian2022} by means of numerical simulations.

The quantum simulation of the chiral BF theory opens interesting perspectives. First, since our mapping of a Raman-coupled BEC into the encoded chiral BF Hamiltonian remains valid at the quantum level, our approach allows the investigation of the chiral BF theory beyond its classical limit, i.e. exploring chiral BECs beyond the mean-field approximation. Moreover, chiral BECs provide a route to experimentally probe the peculiar quantum many-body bound states of the chiral BF theory \cite{Aglietti1996, Jackiw1997, Griguolo1998}, whose classical limit are the chiral solitons that we have recently observed \cite{Froelian2022}. Second, the chiral BF theory corresponds to a field-theoretical formulation of the Kundu linear anyon model \cite{Kundu1999}. It would thus be interesting to explore its connections with other proposals for realizing linear anyon in optical lattices \cite{Keilmann2011, Greschner2015, Straeter2016}, clarifying the differences existing between continuum and lattice linear anyon models \cite{Bonkhoff2021}. Third, although in this paper we have focused on the chiral BF theory on a line, i.e. on a system with open boundary conditions, the model can also be formulated in a ring, i.e. with closed boundary conditions. In this case, the gauge degrees of freedom cannot be completely eliminated any longer: new topological configurations appear that correspond to the integer values of the magnetic flux piercing the ring. Such self-generated magnetic flux could be realized with state-of-the-art experimental techniques \cite{Edmonds2013}. To this end, our Raman-coupling experimental scheme should be applied to a BEC trapped in an annular geometry \cite{Ramanathan2011, Moulder2012, Corman2014}, and extended to a beam configuration that imparts angular momentum in a so-called spin-orbital angular momentum coupling scheme \cite{Chen2018a, Chen2018b, Zhang2019}. Finally, using a position-space treatment in the regime of large Rabi frequency, Ref. \cite{Valenti2020} proposes to implement Chern-Simons theory by applying such spin-orbital angular momentum coupling scheme to a two-dimensional system. Our momentum-space microscopic treatment of interactions in the Raman-coupled system should provide a deeper insight into the meaning of this mapping and allow us to assess its validity for realistic experimental conditions. Realizing such a paradigmatic topological gauge theory in a neutral-atom quantum simulator would provide a new experimental platform to engineer anyonic excitations, fostering the development of adequate experimental tools to manipulate them and characterize their properties without the need for strong correlations.
\\
\begin{acknowledgements}
We thank C. R. Cabrera, M. Dalmonte, T. Grass, V. Kasper, P. \"{O}hberg and G. Valent\'{i}-Rojas for discussions, and the Quantum Optics theory group at ICFO for providing access to their computer cluster. Research at ICFO was funded by European Union (ERC CoG-101003295 SuperComp), MCIN/AEI/10.13039/501100011033 (LIGAS PID2020-112687GB-C21 and Severo Ochoa CEX2019-000910-S), Deutsche Forschungsgemeinschaft (Research Unit FOR2414, Project No. 277974659), Fundaci\'{o}n Ram\'{o}n Areces (project CODEC), Fundaci\'{o} Cellex, Fundaci\'{o} Mir-Puig, and Generalitat de Catalunya (AGAUR and CERCA program). Research at UAB was funded by MCIN/AEI/10.13039/501100011033 (LIGAS PID2020-112687GB-C22). Both groups acknowledge funding from the QuantERA project DYNAMITE (PCI2022-132919 funded by MCIN/AEI/10.13039/501100011033 and by the European Union NextGenerationEU/PRTR) and from the European Union Regional Development Fund within the ERDF Operational Program of Catalunya (project QUASICAT/QuantumCat Ref. No. 001-P-001644). C. S. C. acknowledges support from the European Union (Marie Sk{\l}odowska-Curie–713729), A. F. from La Caixa Foundation (ID 100010434, PhD fellowship LCF/BQ/DI18/11660040) and the European Union (Marie Sk{\l}odowska-Curie–713673), E. N. from the European Union (Marie Sk{\l}odowska-Curie–101029996 ToPIKS), R. R. from the European Union (Marie Sk{\l}odowska-Curie–101030630 UltraComp), L. T. from MCIN/AEI/10.13039/501100011033 and ESF (RYC-2015-17890), and A. C. from the UAB Talent Research program.
\end{acknowledgements}


%

\end{document}